\documentclass[preprint,12pt]{elsarticle}

\usepackage{graphicx}

\usepackage{epstopdf, epsfig}
\usepackage{mathtools}
\usepackage{latexsym}
\usepackage{wasysym}
\usepackage{amssymb}
\usepackage{siunitx}
\usepackage[justification=centering]{caption}
\usepackage{lineno}
\usepackage{chngcntr}
\usepackage{hyperref}

\hypersetup{
    colorlinks=true,
    citecolor=black,
    filecolor=black,
    linkcolor=red,
    urlcolor=black
}
\usepackage[nameinlink]{cleveref}
\usepackage{indentfirst}
\usepackage{natbib}

\journal{Applied Acoustics}

\begin{document}

\begin{frontmatter}

\title{The Sound Radiated by Tip Clearances Submerged in a Boundary Layer}

\author[first]{Manuj Awasthi}
\author[second]{Danielle Moreau}
\author[third]{Paul Croaker}
\author[fourth]{Paul Dylejko}

\address[first]{University of New South Wales, Sydney, NSW 2052, Australia}
\address[second]{University of New South Wales, Sydney, NSW 2052, Australia}
\address[third]{Defence Science and Technology Group, Melbourne, VIC 3207, Australia}
\address[fourth]{Defence Science and Technology Group, Melbourne, VIC 3207, Australia}

\begin{abstract}

The present study investigates the behaviour of the far-field sound radiated by low Mach number tip clearance flow induced by placing a stationary cambered airfoil adjacent to a stationary wall. The tip clearance heights ranged from 14\% to 30\% of the incoming, undisturbed boundary layer thickness and the clearance heights based Reynolds numbers were between 2,600 and 16,000. The far-field sound measured using a microphone array was beamformed to reveal the dominant noise sources and how they behave when the flow Mach number, angle of attack and the clearance height were varied. \textcolor{black}{The near-field behaviour was also examined through PIV measurements and surface pressure fluctuation measurements on the tip}. The results show that the mid-to-high frequency noise generated by tip clearances is dominated by the \textcolor{black}{leakage flow in the mid-chord and leading-edge regions, while a distinct low-frequency noise source with a different scaling behaviour exists close to the trailing-edge of the tip clearance. The origin of this low-frequency noise source is believed to be the tip separation vortex that resides close to the trailing-edge and induces significant turbulence levels in the region. The strength of this noise source decreases with clearance height which is consistent with a reduction in turbulence levels associated with the separation vortex. The magnitude of the mid-frequency clearance noise which scales with the sixth power of the Mach number, decreases with tip clearance height due to a reduction in the fluctuating pressure on the airfoil tip surface. The time-scale of this sound was independent of the flow velocity, implying that the source is non-compact. Smaller tip clearances were also found to generate louder high-frequency noise due to intense turbulence and pressure fluctuation levels concentrated near the leading-edge of the clearance.} \par

\end{abstract}

\begin{keyword}
Tip Clearance Noise; Pressure Fluctuations; Aeroacoustics.
\end{keyword}

\end{frontmatter}

%\linenumbers

\section{Introduction}

The flow between the tip of an airfoil and an adjacent wall, termed tip clearance flow or tip leakage flow ($TLF$), has historically received significant interest due to the pressure losses it induces in turbomachinery applications such as flow through axial compressors, turbines and ducted fans. Thus, there have been a number of studies which have investigated the tip clearance flow-field in axial rotors \cite{RN259,RN260,RN261,RN262,RN258} or in a linear cascade setup with both stationary \cite{RN255,RN265} and moving walls \cite{RN256,RN266,RN264}. Regardless of the configuration, the tip clearance flow-field is known to be complex and three-dimensional, and consists of multiple vortex systems, anisotropic boundary layers and separated flow. Furthermore, the behaviour of these flow features are a function of the blade loading (angle of attack, $\alpha_g$), the tip clearance height ($h$) and the character of the approaching boundary layer on the adjacent wall. While the flow-field around the tip clearance has been studied in great detail historically, the sound radiated by tip clearance flows has received significantly less attention. Although, there have been a few studies \cite{RN268,RN270,RN269} which have measured (and even attempted to reduce) noise from tip clearance flows in turbomachinery, the relationship between the sound and its aerodynamic origins is not well understood due to difficulties associated with isolating the clearance noise from other sound sources. Additionally, the effect of changing the angle of attack and the tip clearance height relative to the incoming boundary layer thickness ($\delta$) on the sound source is not well documented. In light of these facts, the present study is dedicated to understanding the behaviour of the sound radiated by an idealised tip clearance geometry -- a single stationary airfoil adjacent to a stationary wall. \par

An idealised tip clearance geometry and associated flow is a simplification of the more complex clearance flows found in turbomachinery, but it allows a study of the fundamental aeroacoustic sources in $TLF$ by excluding the effects of other turbomachinery flow features such as centrifugal forces, blade-to-blade interactions and the relative motion between the blade and the casing. Moreover, the idealised nature of this configuration permits a localisation of aeroacoustic sources associated with the $TLF$ without contamination from other sources in conventional rotating machinery rigs (e.g. tonal blade passage noise, rotor wake-stator interaction noise, motor noise etc.). The idealised tip clearance configuration has recently been utilised by several authors \cite{RN252,RN253,RN251,RN254} to study the flow-field and sound radiation from tip clearance flows. \par

To understand the utility of studying the noise from an idealised tip clearance configuration, one may consider the study by Ganz \textit{et al.} \cite{RN270} who measured the broadband noise from a fan-rig and isolated the rotor-boundary layer interaction noise by performing separate measurements with and without the casing boundary layer present, and then subtracting the spectra. Despite this, the isolated sound levels include noise from both the $TLF$ and any interaction between the blades and the casing boundary layer makes it difficult to isolate the underlying noise generation mechanisms. Thus, an understanding of the aeroacoustics of an idealised $TLF$ can aid in more informed interpretation of turbomachinery noise data. Furthermore, the knowledge of the behaviour of the sound sources associated with $TLF$ can also help develop better noise control strategies. \par

The salient feature of the $TLF$ is the cross flow from the pressure to the suction side of the blade which eventually rolls up into a tip leakage vortex ($TLV$). The $TLV$ is known to be the dominant source of unsteadiness in $TLF$ and has been studied widely since it can induce significant aerodynamic losses as it interacts with the blade wake downstream as well as with the flow field of the adjacent blades. The behaviour of the $TLV$ both in the vicinity of the blade and further downstream has been described in detail by several authors (see \cite{RN261,RN255}, for example). Aeroacoustically, the cross flow within the tip clearance is of interest since the pressure fluctuations interact with surface discontinuities on the suction and pressure sides of the blade and are scattered to the far-field as sound. Based on the conditional averaging of particle image velocimetry ($PIV$) datasets (originally measured by Grilliat \textit{et al.} \cite{RN189}) trigerred by events in the far-field sound measurements, Camussi \textit{et al.} \cite{RN272,RN273} have shown that for a stationary blade with $h/\delta$ = 0.55 and $\alpha_g$ = 15$^\circ$ the dominant sound source is located within the tip clearance gap, between approximately 40\% to 60\% chord location. Their results, however, do not provide a breakdown of the clearance noise sources according to frequency and other geometric parameters such as angle of attack and the tip clearance height. The breakdown of sound sources according to frequency is important if one considers the results of Koch \textit{et al.} \cite{RN253} ($h/\delta$ = 1.42, $\alpha_g$ = 15$^\circ$) who have shown that while the mid-chord region where the $TLF$ resides is indeed the dominant sound source for $fc/U_\infty >$ 5.7 (where $f$ is frequency in Hz, $c$ is the airfoil chord-length and $U_\infty$ is the free-stream velocity), at lower frequencies the sound from the trailing-edge region is comparable to that from the mid-chord region. This  noise from the trailing-edge region of the tip could be a result of the presence of a tip separation vortex ($TSV$) which has been observed in both idealised clearance flows \cite{RN253} and linear cascade studies with both stationary \cite{RN265} and moving walls \cite{RN266}. The $TSV$ forms due to flow separation inside the gap and although it is not as prominent a flow feature as the well-studied $TLV$, it is important aeroacoustically since it resides near the sharp edges of the tip where the unsteadiness produced by its presence can be scattered efficiently to the far-field. \par

The angle of attack can have a significant influence on the radiated sound as it dictates the behaviour of both the $TLF$ and the $TLV$. With increasing angle of attack, the $TLF$ becomes stronger and moves further upstream. The stronger cross-flow means a stronger flow separation from the tip surface, as well as a more intense $TLV$. These two factors can lead to significant turbulence generation and ultimately an increase in radiated sound levels. Grilliat \textit{et al.} \cite{RN189} who performed PIV measurements in the wall-parallel plane inside a tip clearance formed by a stationary NACA 5510 blade for two geometric angles of attack ($\alpha_g$) of 5$^\circ$ and 18$^\circ$ have shown that an increase in angle of attack leads to a substantial increase in the magnitude of the fluctuating component of the cross-flow velocity. They have shown that an increase in angle of attack not only enhances the unsteadiness associated with the $TLV$, it also increases velocity fluctuations closer to the suction side edge of the tip between the mid-chord and the trailing-edge of the clearance. A similar region of unsteadiness can be observed in simulations of Koch \textit{et al.} \cite{RN253} and could be a result of the presence of flow separation and the $TSV$ in this region. One can presume that, in addition to a stronger cross-flow, such high turbulence levels associated with the $TSV$ can also generate more noise as the angle of attack increases. However, at present, the location of important sound sources in tip clearance flows, their frequency content and how they behave with a change in angle of attack is not known. \par

Besides the angle of attack, the tip clearance height is another important parameter which can affect the sound generation significantly. An increase in the tip clearance height promotes the formation of stronger cross flow and $TLV$s, and larger tip clearances are known to produce more sound in the case of ducted axial fans \cite{RN267}. On the other hand, smaller tip clearances are known to shift the $TLF$ further upstream \cite{RN261} and in case of an idealised tip clearance flow, they can increase the velocity fluctuations towards the suction-side of the tip near the trailing-edge as evident in PIV results of Grilliat \textit{et al.} \cite{RN189}. Ganz \textit{et al.} \cite{RN270} who measured noise from an axial ducted fan have also shown that the far-field sound is sensitive to tip clearance height, particularly for low angles of attack where smaller tip clearances show a strong high-frequency ($>$ 10 kHz) noise component in the far-field. As they derived this result by linearly subtracting the rotor noise without boundary layer from the noise with the boundary layer present, it remains unclear whether these differences are due to rotor-boundary layer interaction, or due to modification of the tip clearance flow by the boundary layer. The complex nature of the far-field sound on the tip clearance height in an idealised clearance flow has recently been demonstrated by Palleja-Cabre \textit{et al.} \cite{RN251} who have revealed that the effect of lowering the tip clearance height on the far-field sound is frequency-dependent. They found that the noise for $fc/U_\infty <$ 24.4 dropped rapidly once the gap height was reduced below 10 mm, while the high frequency noise ($fc/U_\infty >$ 48.8) remained nearly independent of the tip clearance height. Interestingly, Ganz \textit{et al.} \cite{RN270} found the high-frequency sound to be dependent upon clearance height (as well as angle of attack) which could suggest that the noise increase in their measurements was related to rotor-boundary layer interaction since their study was performed on a ducted fan, rather than an idealised tip clearance flow. Alternatively, the difference in the character of the incoming boundary layer and the relative sizes of the tip clearances could also be responsible for these differences. Regardless, these conflicting results highlight the need for a systematic study of the behaviour of the far-field sound and the dominant source regions so that the tip clearance flow effects can be distinguished clearly from the rotary effects. \par

Here, the study of an idealised tip clearance flow where a single, stationary airfoil blade is placed adjacent to a stationary, flat wall is undertaken to better understand the aeroacoustics of tip leakage flow. This geometry precludes any noise generated due to blade-wall relative motion, centrifugal force effects, blade-to-blade interactions, or rotor-stator interactions. \textcolor{black}{Additionally, the airfoil profile used in the present work (a variant of NACA 65A-010 profile) has a smaller camber (and lift) than that found in typical turbomachinery applications. Despite these differences, several tip clearance flow features and the noise generation mechanisms that exist in this ideal configuration (e.g., tip leakage and separation vortex systems and flow separation from the tip edges) are expected to be similar to those found in more practical configurations, and therefore the results of this study are relevant to real-world tip clearance flows.} The idealised tip clearance flow configuration is similar to several recent studies \cite{RN189,RN272,RN273,RN274,RN252,RN254,RN251,RN253} which have attempted to understand (and even reduce) the noise generated from an isolated tip clearance flow. The present study expands upon these studies by localising the dominant sound sources as a function of frequency, angle of attack and tip clearance height. Furthermore, broadband measurements of surface pressure fluctuations have been made at several locations on the tip surface to further understand the behaviour of the source region. Three tip clearance heights equal to 1.3\%, 2.0\% and 2.6\% of the tip chord-length were considered in the present work. Each tip clearance considered is smaller than the height of the incoming, undisturbed boundary layer with clearance ranging between 14\% and 30\% of the approaching, undisturbed boundary layer height. The smallest tip clearance height considered in the present study is of the order of the momentum thickness of the undisturbed boundary layer and the Reynolds numbers (based on clearance height) range between 2,600 to 16,000. The Mach number ($M_\infty$) in the present study ranged from 0.04 to 0.13 which corresponds to free-stream velocities between 15 m/s and 45 m/s, respectively. \par

\section{Experimental Setup and Instrumentation}

\subsection{The UNSW Anechoic Wind Tunnel (UAT)} \label{sec:UAT}

\begin{figure}
\centering
  \centerline{\includegraphics{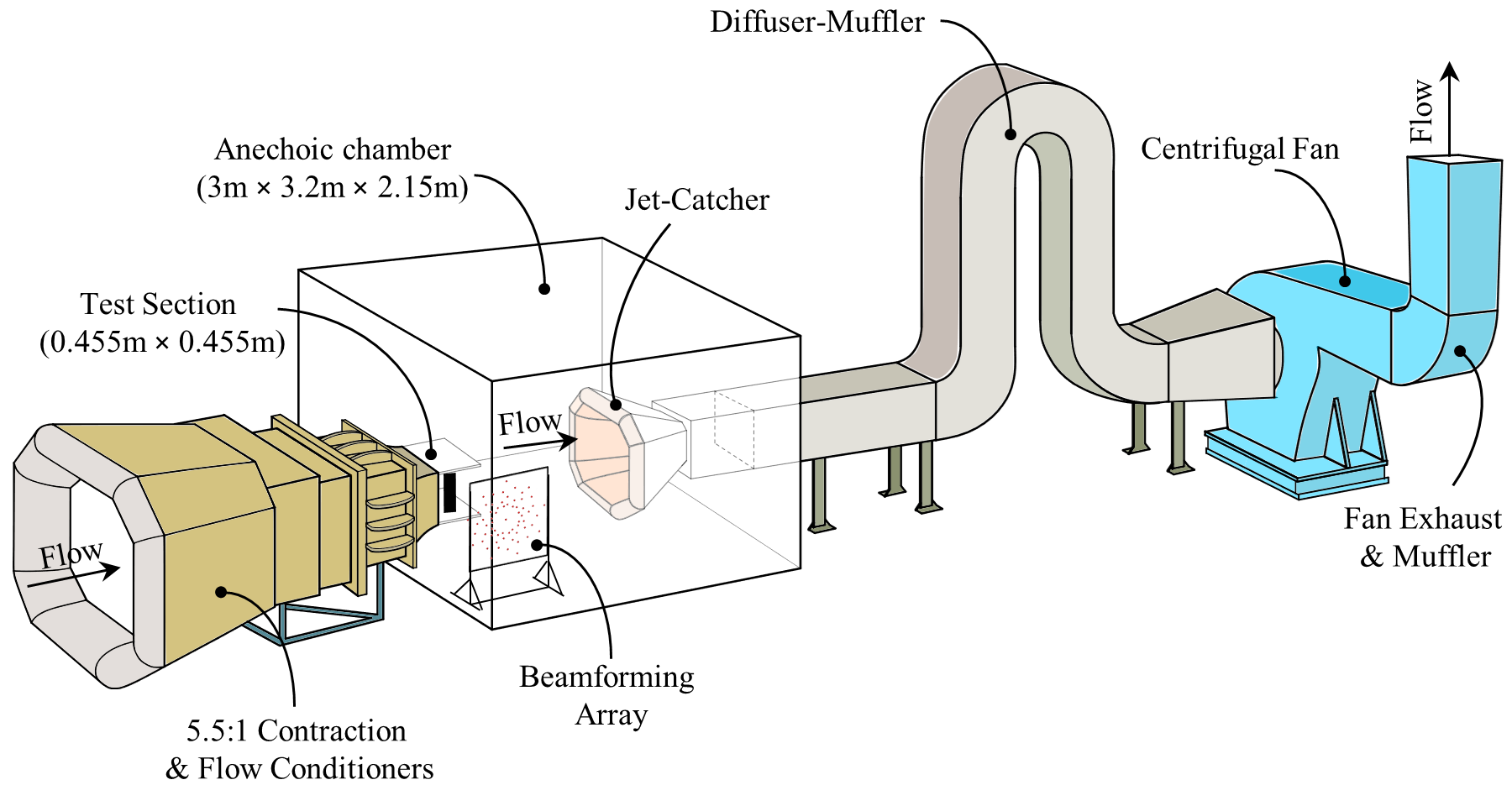}}
  \caption{\textcolor{black}{The UNSW Anechoic Wind Tunnel (UAT) and microphone array. Figure adapted from \cite{RN334}.}}
\label{fig:UAT_Sketch}
\end{figure}

The measurements were performed in the University of New South Wales Anechoic Wind Tunnel (UAT) (\textcolor{black}{see \cref{fig:UAT_Sketch} below}) which is an open-jet type wind tunnel facility with a 0.455 m $\times$ 0.455 m test-section surrounded by a 3 $\mathrm{m}$ × 3.2 $\mathrm{m}$ × 2.15 $\mathrm{m}$ anechoic chamber, see \cite{RN334} for a detailed description of the facility. The facility is driven by an AirEng\textsuperscript{\textregistered} centrifugal fan (volumetric flow rate of 14.2 $\mathrm{m^3/s}$ against a pressure drop of 10 $\mathrm{kPa}$) which is placed at the downstream end of the facility. The fan is connected to the anechoic chamber through an acoustically-lined duct with a U-bend which serves both as a diffuser to expand the flow, and as a muffler to attenuate the fan noise travelling upstream to the chamber. The walls of the anechoic chamber are acoustically treated with 0.25 m thick Melamine foam, creating an anechoic environment above approximately 350 $\mathrm{Hz}$. The floor of the chamber consists of 50 $\mathrm{mm}$ thick perforated plates backed by acoustic-grade foam to minimise reflections of the test model noise from the floor. At the upstream end of the facility, a honeycomb inlet with five turbulence reduction screens condition the flow and a 5.5:1 contraction serves to accelerate the flow to test velocities. The flow enters the anechoic chamber through a 0.455 $\mathrm{m}$ $\times$ 0.455 $\mathrm{m}$ inlet and forms an open-jet type flow. The jet entering the anechoic chamber through the inlet is captured by an acoustically lined, bell-mouth jet-catcher placed 1.7 $\mathrm{m}$ downstream of the inlet, resulting in a 0.455 $\mathrm{m}$ $\times$ 0.455 $\mathrm{m}$ $\times$ 1.7 $\mathrm{m}$ test-section. Far-field acoustic measurements are performed by placing instrumentation (microphone or microphone array) inside the chamber, but outside the flow region \textcolor{black}{as shown in \cref{fig:UAT_Sketch}.} \par

In the present work two parallel endplates were bolted to the test-section inlet to create a tip clearance flow. The undisturbed (no airfoil present) boundary-layer velocity profiles were measured 18 mm downstream of the airfoil leading-edge location (in absence of the airfoil) using a single-sensor hotwire probe (TSI model 1218) connected to a Dantec Dynamics streamware 90N10 constant temperature anemometer. The hotwire probe had a prong spacing of 1.25 mm and used a \SI{5}{\um} platinum-plated tungsten wire as the sensor wire. The hotwire was calibrated for any temperature drifts before and after each boundary layer sweep, and the calibrations were applied to the anemometer output voltages using King's law. The undisturbed boundary layer parameters at three different velocities are shown in \cref{tab:undisturbed_bl_parameters}. \textcolor{black}{Note that, although the parameters are measured 18\,mm downstream of the airfoil leading-edge location due to structural limitations, they are representative of the incoming flow characteristics since any differences between these and those near the airfoil leading-edge would be minimal, owing to the zero pressure gradient in the test section.}

\begin{figure}
\centering
  \centerline{\includegraphics{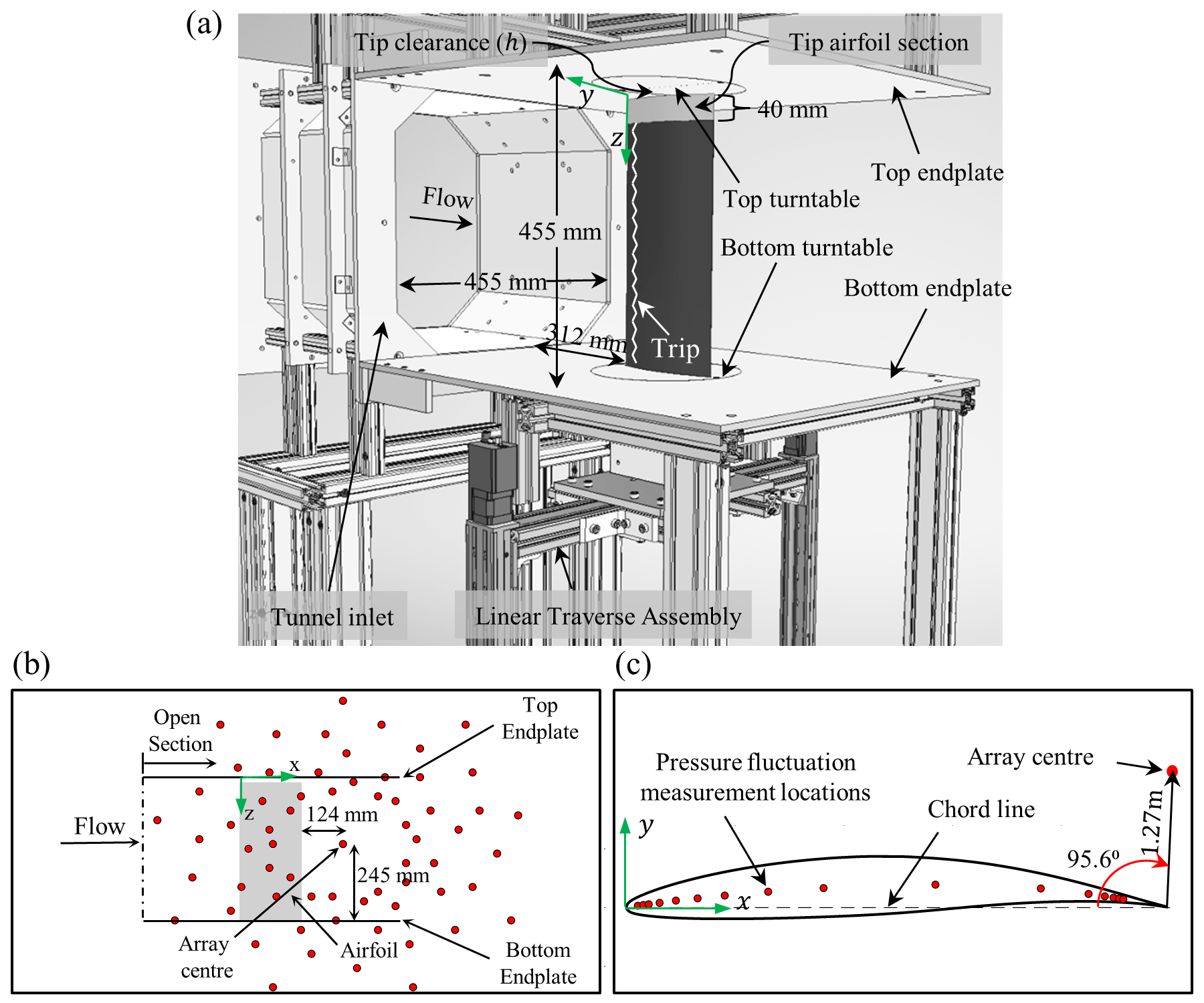}}
  \caption{\textcolor{black}{(a) The tip clearance flow test rig installed in the UNSW Anechoic Wind Tunnel (UAT). The far-field microphone array location relative to the airfoil in the $x-z$ and $x-y$ planes are shown in (b) and (c), respectively. The locations of surface pressure fluctuation measurements (discussed in \cref{sec:Surf Pressure Instrumentation}) on the tip surface are also shown in (c).}}
\label{fig:Tip_Gap_Test_Rig}
\end{figure}

\begin{table}
  \begin{center}
\def~{\hphantom{0}}
 \caption{Undisturbed boundary-layer parameters on the top endplate 18\,mm downstream of the airfoil leading-edge location.}
\resizebox{\textwidth}{!}{\begin{tabular}{ccccccccc}
\\[-1.8ex]\hline \\
$U_\infty$ (m/s) & $\delta$ (mm) & $\theta$ (mm) & $\delta^*$ (mm) & $H$ & $u_\tau$ (m/s) & $Re_\delta$ & $Re_\theta$ & $Re_\tau$ \\
\\[-1.8ex]\hline \\        
15 m/s & 18.8 & 2.22 & 3.09 & 1.39 & 0.64 & 18,279 & 2,152 & 774 \\
30 m/s & 17.8 & 2.01 & 2.70 & 1.35 & 1.22 & 34,651 & 3,902 & 1,392  \\
45 m/s & 17.9 & 1.90 & 2.49 & 1.31 & 1.76 & 51,278 & 5,422 & 2,022  \\
\\[-1.8ex]\hline 
\end{tabular}}
  \label{tab:undisturbed_bl_parameters}
  \end{center}
\end{table}

\subsection{Tip Clearance Test Rig} \label{sec:Test Setup Description}

A purpose-built, test rig (shown in \cref{fig:Tip_Gap_Test_Rig} (a)) capable of traversing an airfoil in a direction perpendicular to a wall (to vary the tip clearance) was designed and installed at the inlet of the test-section. The test-rig consisted of two parallel, 825 mm $\times$ 590 mm $\times$ 9.6 mm endplates which were bolted to the facility inlet and the tip clearance height ($h$) is the distance between the airfoil tip and the top endplate. Each endplate contained a 240 mm diameter turntable capable of being rotated and locked to any desired angle of attack with respect to the flow. To vary the tip clearance height, the bottom turntable consisted of a slot through which the airfoil could be moved freely in a direction perpendicular to the endplate. The airfoil was split into two spanwise segments -- a 500 mm span base airfoil section and a 40 mm span tip airfoil section. The two sections were joined together using countersunk bolts on the tip surface. Once bolted together, the countersunk holes were filled using an automotive filler and sanded to ensure there was no discontinuity on the tip surface. The airfoil section protruded beneath the bottom turntable and was bolted to a linear traverse assembly built by pairing two Zaber X-LRQ450BL-E01\textsuperscript{\textregistered} linear actuators which have a positional accuracy of \SI{60}{\um}. This setup allowed a motorised movement of the airfoil to obtain tip clearance heights of $h$ = 5.28 mm, 3.96 mm, and 2.64 mm. The initial position of the tip surface relative to the top endplate was obtained by using a laser proximeter with an uncertainty of $\pm$ 0.05 mm which is then the uncertainty in the tip clearance height. \par

The airfoil section used in the present work was a cambered airfoil derived by modifying a NACA 65A-010 profile, \textcolor{black}{see \cref{fig:Tip_Gap_Test_Rig} (c)}. The airfoil has a closed trailing-edge with a theoretical chord-length of 200 $\mathrm{mm}$. For machining purposes, the actual airfoil chord-length was truncated to 198.5 $\mathrm{mm}$ which yields a 0.1 $\mathrm{mm}$ thick trailing-edge. The sharp trailing-edge minimises the shedding noise generated by bluff trailing edges. The airfoil has a 4\% camber with the maximum camber located at the 60\% chord location. In the present work, the theoretical chord-length ($c$ = 200 mm) will be used as the reference length. \Cref{tab:Airfoil and tip clearance parameters} shows the chord-based Reynolds numbers, tip clearance based Reynolds numbers and the tip clearance height in terms of the approaching undisturbed boundary layer and momentum thicknesses. The range of tip clearance heights tested were approximately between 14\% and 30\% of the incoming boundary layer thickness and the height of the smallest clearance is of the order of the undisturbed boundary layer momentum thickness. Since the Reynolds number range within the present work lies in the regime where the tonal noise due to laminar boundary layer separation on the airfoil surface is prevalent \cite{RN16,RN17}, the airfoil boundary layer was tripped at the 10\% chord location using a 0.4 mm tall zig-zag tape as illustrated in \cref{fig:Tip_Gap_Test_Rig}. The trip tape was terminated 40 mm below the tip to avoid the contamination of the leakage flow by the trip. A comparison of the sound radiated with and without the tape showed that the artificial tripping successfully suppresses the tonal noise associated with the laminar boundary layer separation. The surface pressure fluctuations measured on the tip surface (discussed later in \cref{sec:Surf Pressure Instrumentation}) with and without the trip were found to be identical, suggesting that the trip does not affect the clearance flow. \par  

The coordinate system which will be followed throughout the text is shown in \cref{fig:Tip_Gap_Test_Rig}. The origin of the coordinate system is located on the wall and coincides with the leading edge of the airfoil tip, and the chord-wise, chord-normal and wall-normal coordinates are represented by $x$, $y$ and $z$, respectively. The airfoil was rotated about a point located along the camber line at $x/c$ = 0.45 to obtain three geometric angles of attack, $\alpha_g$ = 5.6$^\circ$, 9.1$^\circ$ and 12.5$^\circ$. The uncertainty in the angle of attack was 0.15$^\circ$. \par

\begin{table}
  \begin{center}
\def~{\hphantom{0}}
 \caption{Tip clearance parameters and Reynolds Numbers}

\begin{tabular}{cccccccc}
\\[-1.8ex]\hline \\
$U_\infty$ (m/s) & $h (mm)$ & $Re_c$ & $\delta (mm)$ & $\theta$ (mm) & $Re_h$ & $h/\delta$ & $h/\theta$ \\
\\[-1.8ex]\hline \\     

15 m/s & 2.64 & 198,625 & 18.8 & 2.22 & 2,621 & 0.14 & 1.19 \\
 & 3.96 &  & & & 3,933 & 0.21 & 1.78 \\
 & 5.28 &  & & & 5,244 & 0.28 & 1.19 \\
30 m/s & 2.64 & 397,250 & 17.8 & 2.01 & 5,244 & 0.15 & 1.30 \\
 & 3.96 &  & & & 7,866 & 0.22 & 1.97 \\
 & 5.28 &  & & & 10,487 & 0.30 & 2.62 \\
45 m/s & 2.64 & 595,875 & 17.9 & 1.90 & 7,866 & 0.15 & 1.39 \\
 & 3.96 &  & & & 11,798 & 0.22 & 2.08 \\
 & 5.28 &  & & & 15,731 & 0.30 & 2.78 \\

\\[-1.8ex]\hline 
\end{tabular}
  \label{tab:Airfoil and tip clearance parameters}
  \end{center}
\end{table}  

\subsection{Far-Field Sound Measurements} \label{sec:Far Field Sound Instrumentation Description}

The far-field sound was measured using a 64-channel phased microphone array placed on the suction-side of the airfoil. \textcolor{black}{The distribution of the array microphones in the $x-z$ plane relative to the airfoil is shown in \cref{fig:Tip_Gap_Test_Rig} (b), while the location of the array centre (the reference location for results presented later) in the $x-y$ plane relative to the airfoil is shown in \cref{fig:Tip_Gap_Test_Rig} (c).} The \textcolor{black}{0.9\,m diameter array} utilises 64 phase-matched GRAS 40PH-10 CCP free-field array microphones arranged in a spiral pattern. Each microphone had a flat frequency response (to within $\pm$ 2 dB) up to 20 kHz and a nominal sensitivity of 50 mV/Pa. The pressure signals generated by the array microphones were acquired using a National Instrument PXI\textsuperscript{\textregistered} data acquisition system at a sampling rate of 65,536 Hz for 32 seconds. The resulting time-series pressure data were Fourier transformed using the periodogram method to yield the cross-spectral matrix (CSM) which was then beamformed (as explained below) to reveal the acoustic sources. The periodogram was calculated by dividing the time-series into blocks with 8192 samples per block. The blocks were overlapped by 50\% to generate additional records for averaging and a Hanning window was applied to each block to reduce the spectral leakage. The resulting narrowband spectra and cross-spectra had a bandwidth of 8 Hz. The far-field sound was measured for seven different free-stream velocities between 15 $\leq U_\infty \leq$ 45 m/s with an increment of 5 m/s. \par

The narrowband CSM was combined in 1/3\textsuperscript{rd} octave bands and then beamformed using the frequency-domain, conventional sum and delay beamforming approach as explained in Brooks and Humphreys \cite{RN111}. In this approach, the measured CSM is mathematically \textit{steered} to different points in a plane of interest (referred to as the scanning grid) and the output experiences constructive interference where a source is present, revealing the source location. The steering vector formulation used to steer the CSM in the present work is the type II formulation discussed in Sarradj \cite{RN230}. The scanning grid (the plane with sources of interest) was chosen as the plane in which the chord-line of the airfoil lies when the airfoil is placed at $\alpha_g$ = 0$^\circ$. The scanning grid was 2 m $\times$ 2 m with a resolution of 4 mm in both directions. To minimise the background noise contamination of beamforming results, the background CSM (obtained by measuring noise in absence of the airfoil) was subtracted from the CSM obtained in the presence of the airfoil before beamforming. The diagonal of the CSM was also removed before beamforming to reduce the effect of microphone self-noise on the source maps. Unless otherwise noted, the source levels will be presented as 1/3\textsuperscript{rd} octave-band sound pressure levels ($SPL$) normalised on a reference pressure of \SI[prefixes-as-symbols]{20}{\micro\pascal}. Assuming that the dominant uncertainty in the beamformed source levels comes from the uncertainty in the background-subtracted spectral estimates in 1/3\textsuperscript{rd} octave bands, the uncertainty was calculated based on the method outlined in \cite{RN37} and was found to be 1\,dB or less across the frequency range considered. \textcolor{black}{Further note that, as shown in \cref{fig:Tip_Gap_Test_Rig} (b), some of the array microphones lie above and below the endplates and are therefore not in direct line-of-sight of the tip clearance. A comparison of the beamforming output between the full array and a 45-microphone subarray formed by including only the microphones within the direct line-of-sight showed negligible difference (up to 1\,dB). Similar comparisons were obtained when an even smaller array consisting of 30 microphones concentrated near the array centre were utilised for beamforming.} The acoustic waves travelling from the airfoil to the phased array which is located outside the flow region are refracted by the shear-layer present at the edge of the open-jet flow. The primary effect of this refraction is that the sources appear shifted slightly downstream of their actual locations. Padois \textit{et al.} \cite{RN231} have shown that for low Mach number flows as in the present work, this downstream shift can simply be corrected by moving each source upstream by a distance equal to the Mach number times the half-width of the potential core of the jet. This is the approach which was taken to account for the shear-layer refraction effects in the present work. \par

\begin{figure}
\centering
  \centerline{\includegraphics{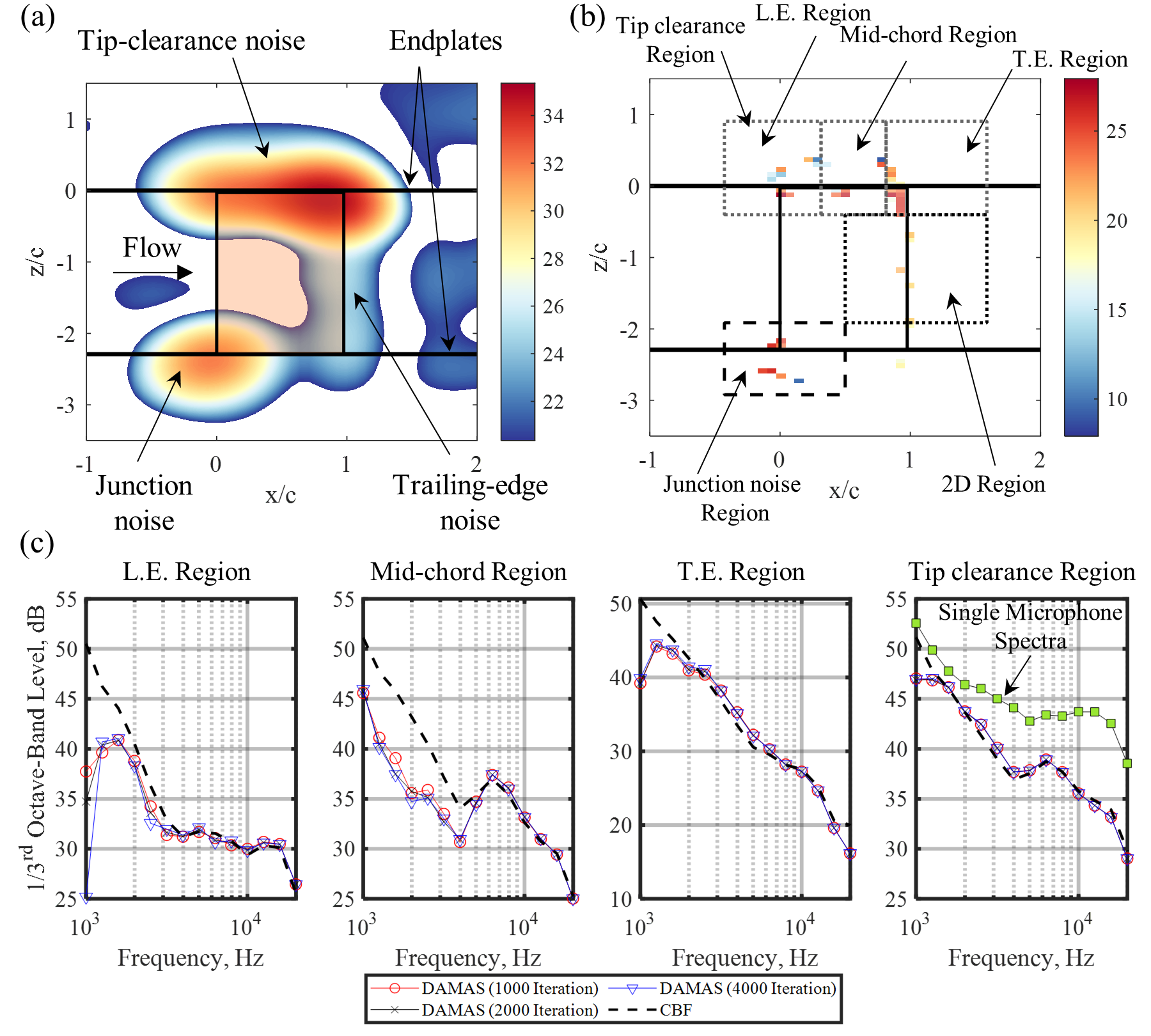}}
  \caption{Tip clearance noise source levels (1/3\textsuperscript{rd} octave-band SPL, $p_{ref}$ = 20$\mu$Pa) at 4 kHz obtained using (a) Conventional beamforming and (b) DAMAS, for $h$ = 5.28\,mm at $\alpha_g$ = 12.5$^\circ$ and $U_\infty$ = 30\,m/s. The dominant noise sources along with the superimposed airfoil and endplates have been highlighted in (a). The sub-regions selected for integrating the DAMAS output have been superimposed on the map in (b). A comparison between the SPL levels for the overall clearance noise and the three clearance sub-regions obtained by integrating CBF source maps and DAMAS source maps with different iteration limits are shown in (c). \textcolor{black}{The rightmost subfigure in (c) also shows the 1/3$^{rd}$ octave-band spectra measured by the single microphone at the array centre.}}
\label{fig:CBF_vs_DAMAS}
\end{figure}

\Cref{fig:CBF_vs_DAMAS} (a) shows a typical tip clearance sound source map obtained using conventional beamforming at 4 kHz for $h$ = 5.28 mm at $\alpha_g$ = 12.5$^\circ$ and $U_\infty$ = 30 m/s. The projection of the airfoil in the chordwise plane at $\alpha_g$ = 0$^\circ$ has been superimposed upon this map along with the parallel endplates. The dynamic range of the colour scale on the map is 15\,dB. Three distinct noise sources -- the tip clearance noise, the airfoil trailing-edge noise and the airfoil-bottom wall junction noise -- can be clearly observed. The tip clearance source is the dominant noise source for this particular case. Although not shown, this is also true for all the beamformed source maps where the airfoil noise levels exceed the background noise levels. Although conventional beamforming source maps reveal the noise sources in the flow, there is always an ambiguity regarding the source location (particularly at lower frequencies) since the beamforming output is convolved with the array point spread function (PSF). This presents a challenge isolating the noise source of interest, particularly in cases where two sources are in close proximity to each other (e.g. the tip clearance and the trailing-edge noise sources in \cref{fig:CBF_vs_DAMAS} (a)). Several methods of minimising the array PSF convolution exist, but in this work the iterative Deconvolution Approach for the Mapping of Acoustic Sources (DAMAS) algorithm of Brooks and Humphreys \cite{RN224} was implemented. \Cref{fig:CBF_vs_DAMAS} (b) shows the source map of (a) after implementation of the DAMAS algorithm. Since the algorithm is computationally expensive, it was implemented on a reduced scanning grid with a spatial resolution of 13.9 mm in both the streamwise and vertical directions, and 2000 iterations were found to be sufficient to obtain convergence. The DAMAS map shows separate clearance noise sources in the leading-edge, mid-chord and trailing-edge regions of the clearance which were not visible in the conventional beamforming map of (a). The DAMAS algorithm, to some extent, also provides a physical separation between the trailing-edge and the tip clearance sources which is not possible in conventional beamforming maps. \textcolor{black}{Note that the DAMAS map shows the presence of several sources above the top endplate. These sources are most likely an artefact of both a limited spatial resolution of the array (point spread function) and the acoustic reflections of the clearance noise by the adjacent endplate. Although, the use of DAMAS minimises the array resolution effects, the reflections which are coherent with the flow noise sources will appear outside of the flow region since the beamforming algorithm utilises a free-field Green's function. In this work, these reflections are considered to be a part of the tip clearance noise, but momentarily we will attempt to quantify the effects of these reflections on the results presented in this paper.} \par

The sample sound source maps presented in \cref{fig:CBF_vs_DAMAS} (a) and (b) are useful in locating sources, but they do not provide an understanding of the sound source behaviour as a function of frequency and other parameters. Thus, in the present work both beamforming and DAMAS source maps were integrated to obtain far-field sound levels. Since the DAMAS source maps provide better separability of sources in the clearance region, the DAMAS source maps were divided into five sub-regions labelled in \cref{fig:CBF_vs_DAMAS} (b) to evaluate the contribution of different source regions to the overall radiated sound. The region labelled \textit{Tip clearance Region} was used to isolate the tip clearance noise, while the regions labelled \textit{2D Region} and \textit{Junction noise Region} were used to investigate the behaviour of the airfoil trailing-edge noise and airfoil - bottom wall junction noise sources, respectively. The \textit{Tip clearance Region} was further divided into three sub-regions as shown. The \textit{L.E. Region} covers the noise sources in the clearance leading-edge region up to the quarter-chord location, the \textit{Mid-chord Region} extends between $x/c$ = 0.25 and 0.75, while the \textit{T.E. Region} includes the sources near the trailing-edge of the clearance. The integrated output of these three regions combined together (\textit{Tip clearance Region}) is taken to be the overall tip clearance noise for each configuration. Each integration sub-region in the clearance region extends approximately 80\,mm below the top wall. This spanwise extent was selected based on visual inspection of several DAMAS maps to ensure that the regions covered the entire clearance source in each region. This presents no issues for the integration output from the \textit{L.E. Region} and \textit{Mid-chord Region} since there are no significant sources nearby, however, the absolute source levels from the \textit{T.E. Region} may always be \textit{contaminated} since it is not possible to clearly separate the airfoil trailing-edge noise from the clearance source in this region. \textcolor{black}{However, as explained below, it is reasonable to assume that the contribution of the airfoil trailing-edge noise to the integrated source levels from the tip clearance region may be negligible.} \par

A sensitivity analysis of the integrated output from the \textit{T.E Region} revealed that shrinking the spanwise extent of this region by 20 mm had no influence on the output, but a further reduction of 20 mm lead to a 2 - 3 dB reduction in the output across the frequency range. This reduction could be either due to exclusion of the trailing-edge noise or it could be due to the shorter region not including the entire clearance noise. A visual inspection of the DAMAS maps does reveal the latter to be true for several frequencies. \textcolor{black}{Another exercise that was undertaken to ensure minimal influence of the airfoil trailing-edge noise on the tip clearance noise spectra was considering the integrated spectra for the three different clearance heights at the same angle of attack as a function of the spanwise extent of the integration region. At the same angle of attack, the airfoil trailing-edge noise should be independent of the tip clearance height. This was indeed found to be the case when the integrated output from the \textit{2D Region} (see \cref{fig:CBF_vs_DAMAS} (b) for the region definition) that includes the airfoil trailing-edge noise was analysed. However, increasing the upper limit of this region even by a pixel showed noticeable variation in the source levels with clearance height for the same angle of attack which suggested that a portion of the clearance noise was being included. This implies that reducing the lower limit of the tip clearance noise integration regions any further will exclude the clearance noise sources, while increasing it any further will include contribution from the airfoil trailing-edge noise. This exercise further showed that the spanwise limits of the integration regions used in the present work are appropriate for extracting the tip clearance noise.} Despite this, the present results from the \textit{T.E Region} must always be interpreted with the caveat that absolute levels may be \textit{contaminated} by the trailing-edge noise. However, given that the clearance noise exceeds the trailing-edge noise substantially, the output is expected to be biased towards the clearance noise. Also note that later, we will evaluate the relative contribution of each sub-region to the overall clearance noise for a fixed angle of attack. In such comparisons, any \textit{contamination} of the \textit{T.E Region} output is not relevant since we are only concerned with relative sound levels. \par

\Cref{fig:CBF_vs_DAMAS} (c) shows a comparison between the integrated tip clearance far-field sound levels obtained by integrating both the conventional beamforming maps and DAMAS maps for $h$ = 5.28 mm and $\alpha_g$ = 12.5$^\circ$ at $U_\infty$ = 30\,m/s. The integrated level comparison is shown for each of the four tip clearance noise regions identified in \cref{fig:CBF_vs_DAMAS} (b). \textcolor{black}{For comparison, the integrated output from the \textit{Tip clearance Region}, which represents the overall tip clearance noise levels, is also compared to the 1/3$^{rd}$ octave-band spectra measured by the single microphone at the array centre.} The integrated output from the conventional beamforming maps was normalised to account for the array PSF effects as outlined by Martinez \textit{et al.} \cite{RN229}. Furthermore, to reduce the side-lobe contamination of the integration output, sources which were 10 dB or lower in magnitude than the maximum source level found in the map at each frequency were ignored in the integration. The integrated levels for DAMAS were obtained for three different iteration limits (1000, 2000 and 4000) to ensure the convergence of the algorithm. \textcolor{black}{We first note that the single microphone spectra (shown in the rightmost subplot) has the same general shape as the integrated CBF and DAMAS levels, but shows elevated levels since it contains contributions from sources outside of the \textit{Tip clearance Region} (airfoil and facility noise source).} The beamforming levels for the overall clearance noise (\textit{Tip clearance Region}) and the \textit{T.E. Region} follow the DAMAS output closely above about 1.5\,kHz, larger differences between the beamformed and deconvolved integration outputs below about 4\,kHz is observed for the source levels in the mid-chord and leading-edge regions. This is because, as shown later, the source levels in these two sub-regions are weaker than those near the trailing-edge and are not well resolved by the beamforming algorithm. The integration output from DAMAS algorithm with different iteration limits are nearly identical (other than the levels in the 1\,kHz bin for the \textit{L.E Region}) and thus all subsequent results shown in this paper were obtained using 2000 iterations of the DAMAS algorithm. Lastly, it is worth noting that the integrated levels for the \textit{Mid-chord Region} \textit{L.E. Region} below about 2\,kHz have larger uncertainty associated with them due to array's poor spatial resolution at such low frequencies and weaker source levels in these regions. \par

\begin{figure}
\centering
  \centerline{\includegraphics{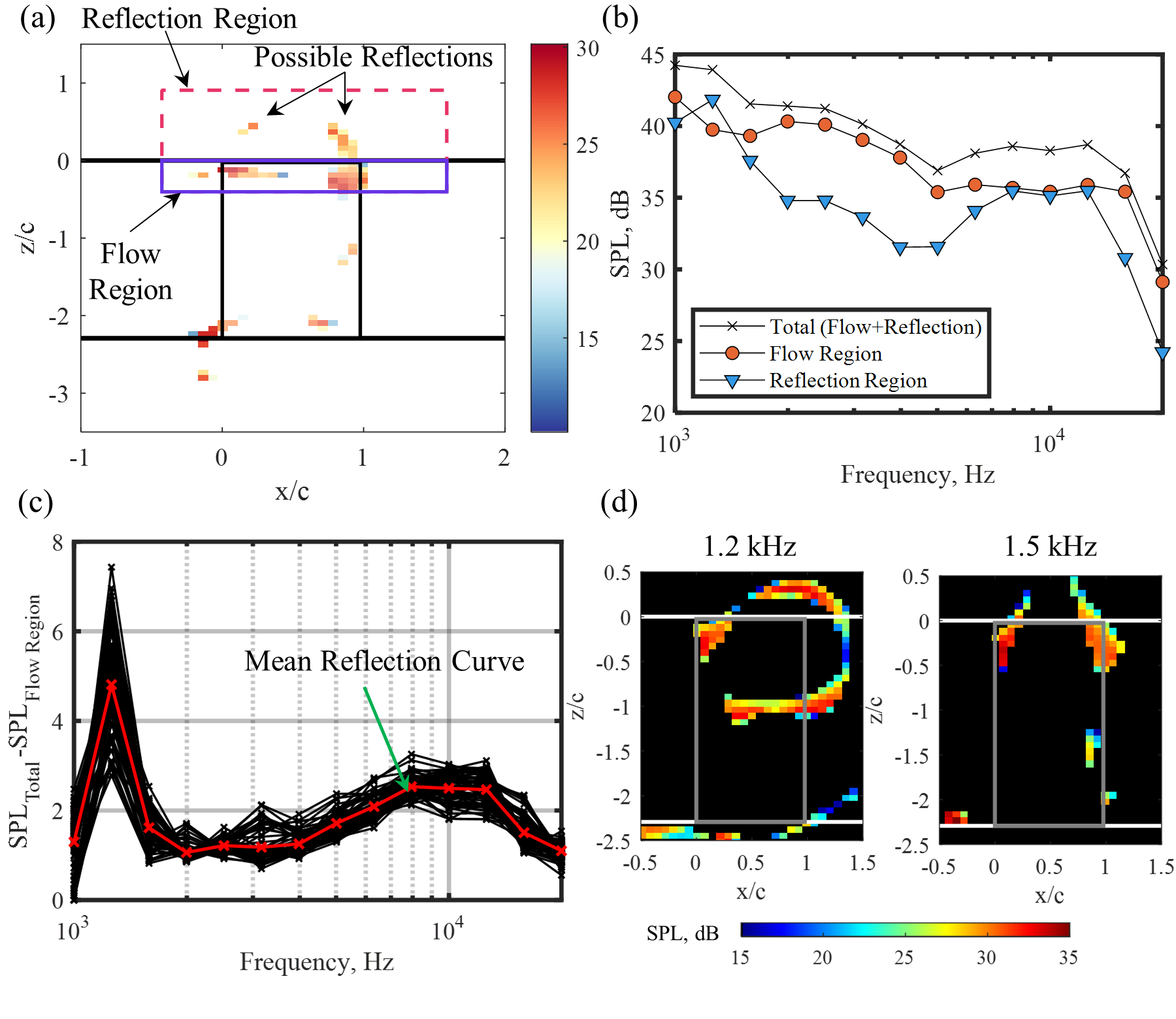}}
  \caption{\textcolor{black}{The effect of reflections on the integrated DAMAS output. (a) The definition of integration regions used to quantify the reflection effects superimposed on the DAMAS sound-source map at 2.5\,kHz for $h$ = 2.64\,mm; $U_\infty$ = 30\,m/s; $\alpha_g$ = 12.5$^\circ$. (b) shows the integrated output for the flow and reflection regions, along with the total integrated levels for the case in (a). The difference between the total clearance noise and the integrated levels from the Flow Region for all the cases considered in the present work (three clearance heights, three angles of attack and seven flow speed) are shown in (c). DAMAS maps at two frequencies (listed above each map) for the case in (a) are shown in (d).}}
\label{fig:Reflection_Effects}
\end{figure}

\textcolor{black}{Finally, in this section we assess the effect of acoustic reflections on the deconvolved beamforming output with the understanding that these reflections themselves are an important part of the tip clearance noise problem. \Cref{fig:Reflection_Effects} (a) shows the DAMAS source map at 2.5\,kHz for $h$ = 2.64\,mm, $\alpha_g$ = 12.5$^\circ$ and $U_\infty$ = 30\,m/s. Superimposed upon this map are the definition of two integration regions that we will utilise to quantify the effects of reflections on the results. The Tip clearance Region depicted in \cref{fig:CBF_vs_DAMAS} (b) has been subdivided into a \textit{Flow Region}, depicted using solid lines, that includes the noise sources in the flow region and a \textit{Reflection Region} that includes the ghost sources due to possible reflections that appear above the endplate. Note that, despite its name, the \textit{Reflection Region} may not exclusively contain acoustic reflections, but also flow noise sources due to limited array resolution, particularly at low frequencies. Nonetheless, this division of integration regions allows an assessment of the ghost source magnitude and how it changes as the flow conditions change. \Cref{fig:Reflection_Effects} (b) shows the integrated acoustic spectra from these two subregions along with the sum of the output from these regions for the same configuration for which the source map is shown in (a). As indicated by these spectra, the strength of the reflections is not uniform across all frequencies with strong reflections observed around 1.2\,kHz and between 8 -- 12\,kHz. It can be shown that these strong reflections are a result of purely acoustic, as opposed to fluid dynamic, phenomena. \Cref{fig:Reflection_Effects} (c) quantifies the effect of reflections on the integrated output for all cases considered in the present work (three clearance heights, three angles of attack and seven flow speeds) by plotting the difference (in dB) between the Total and Flow Region source levels. This plot shows that with the exception of the reflection peak around 1.2\,kHz, the effect of the reflections on the spectra is limited to within 3\,dB of the total source level across the frequency range. Regardless of the geometrical configuration or the flow speed, the reflection magnitude peaks around the same two frequency bands (1.2\,kHz and 8 -- 12\,kHz). The peak in the reflection magnitude around 1.2\,kHz could be associated with either a facility acoustic mode or the inability of the array to provide accurate source localisation at such low frequencies. \Cref{fig:Reflection_Effects} (d) shows DAMAS maps at two frequencies which confirms that the tip clearance noise source around 1.2\,kHz is indeed not resolved accurately. It is also interesting to note that the acoustic wavelength associated with the 1.2\,kHz peak is about 270\,mm which is close to half the distance between the parallel endplates (227.5\,mm) that bound the flow, suggesting that the beamforming results at this frequency could be affected by a coupling between the tip clearance noise and the transverse acoustic duct modes between the two parallel endplates. The wavelengths associated with the higher frequency peaks (8 -- 12\,kHz) in the reflection spectra are much smaller (28\,mm - 42\,mm) and do not correspond to any known geometrical scales, but given their small size they could be associated with acoustic interference effects within the tip gap.}

\subsection{Surface Pressure Fluctuation Measurements} \label{sec:Surf Pressure Instrumentation}

\begin{figure}
\centering
  \centerline{\includegraphics[width = \textwidth]{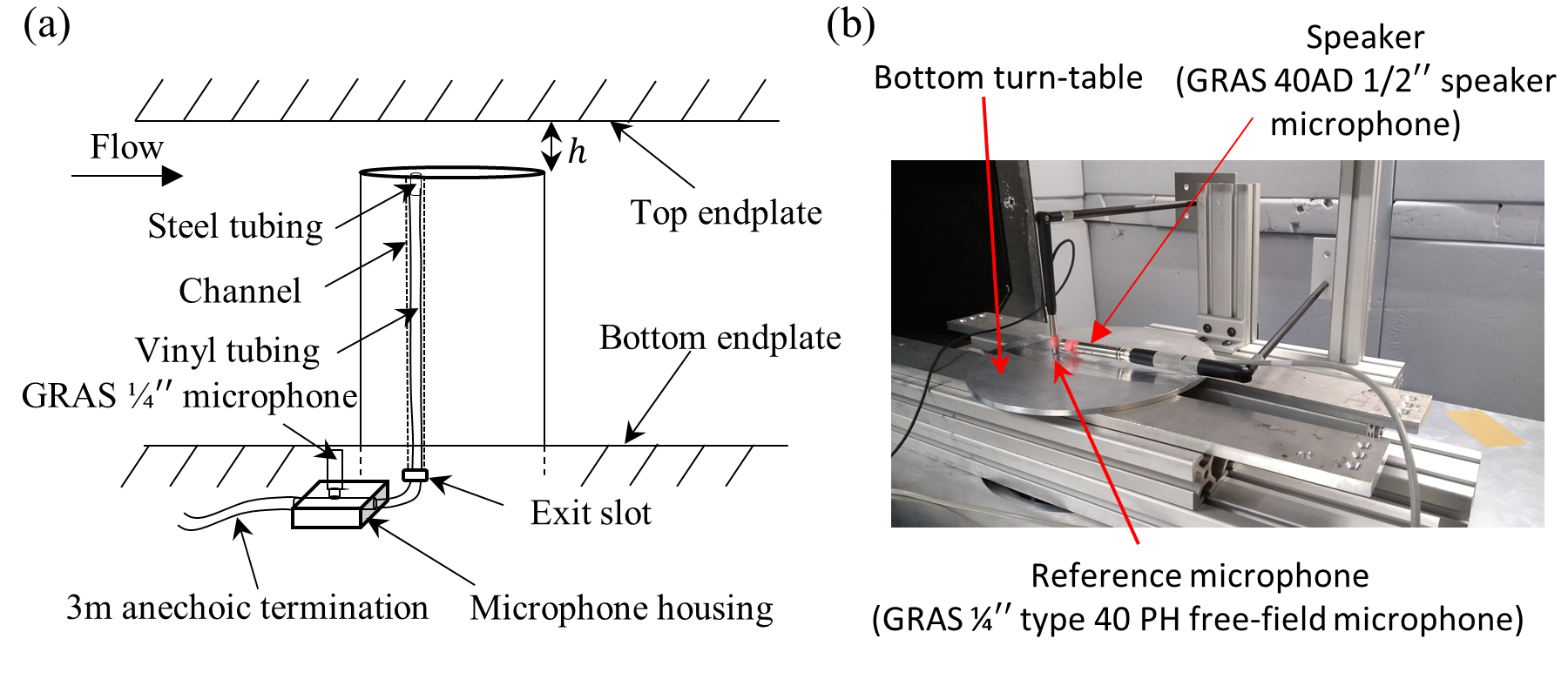}}
  \caption{Remote microphone setup used to measure wall pressure fluctuations on the tip surface. (a) Schematic showing the remote microphone setup (figure is not to scale) and (b) In-situ calibration setup to calibrate remote microphones on the tip surface.}
\label{fig:Remote_Microphone_Setup}
\end{figure}

The surface pressure fluctuations were measured on the airfoil tip surface at 15 different locations distributed along the mean camber line of the airfoil as shown in \cref{fig:Tip_Gap_Test_Rig} inset. \Cref{tab:Surf_Pressure_Locations} lists the chordwise and chord-normal location of these pressure taps. Note that the measurement locations have a finer resolution near the leading and trailing edges compared to those in the mid-chord region. The measurements were made using the remote microphone technique described in Awasthi \textit{et al.} \cite{RN204} which has been shown to yield accurate pressure fluctuations up to about 12 kHz. The remote microphone method \cite{RN204} adapted for measuring the surface pressure fluctuations on the tip surface is illustrated in \cref{fig:Remote_Microphone_Setup} (a) for a single measurement location. To implement the method, 0.8 mm diameter pressure taps were drilled on the tip surface along the camber line. Hollow channels corresponding to each pressure tap were created inside the airfoil by machining the suction and pressure sides of the airfoil separately, creating matching grooves on the mating faces. A 0.5 mm inside diameter steel tubing (nominally 38 mm long) was attached to each tap using epoxy. Care was taken to ensure that the end of this steel tube was flush with the tip surface. A flexible vinyl tubing (1.27 mm and 0.8 mm outside and inside diameter, respectively) was then attached to each steel tubing and tightly sealed to avoid any leakages. The vinyl tubing was then routed through the hollow channels and an exit slot in the airfoil to a 3D-printed module which housed the condenser microphones (GRAS 40 PH-10 1/4 inch microphones; $\pm$2\,dB pressure-field response up to 15 kHz). The housing was placed outside the flow and it consists of a T-junction to route the incoming pressure waves to the microphones and to a 3 m long tubing which serves as an anechoic termination to minimise acoustic reflections within the housing. Once the setup was complete, each pressure tap was checked for leakages by applying a known pressure to each port and measuring the output pressure at the inlet to the microphone housing. \par

The setup depicted in \cref{fig:Remote_Microphone_Setup} (a) allows measurement of surface pressure fluctuations on the tip surface with greater spatio-temporal resolution than the traditional flush-mounted microphone method. It also eliminates the need to mount the microphones within the body of the airfoil. However, the remote microphone setup must be carefully calibrated to account for the pressure losses through the tubing and the housing. The calibrations were performed using the two-step method of Awasthi \textit{et al.} \cite{RN204} which involves comparing the response of each remote microphone to an unmodified condenser microphone (also a GRAS 40 PH-10 1/4 inch microphone) placed in close proximity to each 0.5 mm tap on the tip surface. The unmodified condenser microphone (the reference microphone) is held approximately 2 mm above each pressure tap and both the remote microphone and the reference microphone are then exposed to the same white noise source. The calibration is then formed by dividing the cross-spectrum between the input signal to the speaker and the remote microphone by the cross-spectrum between the input signal and the reference microphone. The two steps in the method -- referred to as a free-field calibration step and a cavity calibration step -- are exactly the same in principle, except in the case of the latter the sound source, the reference microphone, and the pressure tap are enclosed within a cavity. \Cref{fig:Remote_Microphone_Setup} (b) shows an image of the free-field calibration step which includes the sound source and the reference microphone held above a pressure tap on the tip surface. The challenge with calibrating the remote microphone taps on the tip surface is the scattering of noise by the edges of the tip which contaminate the calibration. To overcome this, the bottom turntable (see \cref{fig:Tip_Gap_Test_Rig}), which contains a slot that is the same shape as the airfoil, was mounted flush with the tip surface of the airfoil and any gap between the turntable and the tip surface was sealed with tape. This provides a flat surface around the tip and eliminates the scattering problem allowing each tap to be calibrated. The cavity calibration step (not shown) was performed on the same setup as shown in \cref{fig:Remote_Microphone_Setup} (b), except a hand-held cavity housing the speaker and the reference microphone was placed over each pressure tap.\par

\begin{table}
  \begin{center}
\def~{\hphantom{0}}
 \caption{Tip surface pressure fluctuation measurement locations.}
\resizebox{\textwidth}{!}{\begin{tabular}{cccccccccccccccc}
\\[-1.8ex]\hline \\
$x/c$ (\%) & 2.1 & 3.1 & 4.1 & 6.1 & 9.1 & 18.1 & 26.1 & 36.3 & 56.8 & 76.2 & 85.0 & 88.0 & 89.6 & 90.6 & 91.4 \\
$y/c$ (\%) & 0.3 & 0.4 & 0.6 & 0.8 & 1.2 & 2.0 & 2.7 & 3.3 & 3.9 & 3.2 & 2.3 & 1.9 & 1.7 & 1.5 & 1.4 \\
\\[-1.8ex]\hline 
\end{tabular}}
  \label{tab:Surf_Pressure_Locations}
  \end{center}
\end{table}

The pressure signal from each remote microphone was acquired using a National Instrument PXI\textsuperscript{\textregistered} data acquisition system at a sampling rate of 65,536 Hz for 32 seconds. The pressure time-series was then Fourier transformed using Welch's periodogram method to yield auto and cross-spectral densities between the remote microphones. The parameters used for Welch's methods were the same as those listed for the far-field sound signal processing above. After applying the calibrations, a spectral smoothing was performed by averaging the narrowband spectra over 1/12$^{th}$ octave bands. The surface pressure fluctuations were measured at the same velocities as the far-field sound measurements (15 $\leq U_\infty \leq$ 45 m/s with an increment of 5 m/s). \par

\textcolor{black}{\subsection{Flow-Field Measurements using Particle Image Velocimetry (PIV)} \label{sec:PIV_Setup}}

\textcolor{black}{The flow-field within the gap was measured using 2D PIV in the wall-parallel plane ($x-y$ plane, see \cref{fig:Tip_Gap_Test_Rig} for the coordinate system definition). The measurements were performed for the same three clearance heights and angles of attack as the far-field sound and surface pressure measurements, but at a single velocity of $U_\infty$ = 15\,m/s. A \textit{LaVision} planar 2D-PIV system that utilises two Nd:Yag lasers operating at 15\,Hz synced with two high-resolution CCD cameras was used for the measurements. A Litron NANO-TRL-250-15 Nd:Yag double-pulsed laser was used to generate laser sheets with 250\,mJ/pulse and a wavelength of 532\,nm. The laser sheet was positioned parallel to the wall within the gap such that for each clearance height the sheet centre was aligned with the centre of the clearance height in the wall-normal direction. The flow was seeded using vapourised DEHS particles and two 2048 $\times$ 2048 pixels resolution cameras, each with a 105\,mm lens and placed adjacent to each other, were used to image the particles as they pass through the laser sheet. The time interval between the flow snapshots ($\Delta t$) was 20 $\mu s$ and the raw images from the two cameras were \textit{stitched} together to form a composite image that measured 309.9\,mm and 161\,mm in the $x$ and $y$ directions, respectively.} \par

\textcolor{black}{The raw images were processed using the \textit{LaVision Davis} software to yield the streamwise ($U$) and transverse ($V$) components of velocity. Prior to the PIV processing, the background image obtained using the temporal mean of each dataset was substracted from each snapshot to reduce the reflections from the sharp edges of the tip. The cross-correlation between the seeded flow snapshots was calculated using a multi-pass algorithm with two passes made with a 128 $\times$ 128 pixel window, followed by three passes with a 32 $\times$ 32 pixel window which yields velocity fields in both directions with a resolution of approximately 1.26\,mm in physical space. In each case, the cross-correlation was calculated in the Fourier domain with a 50\% overlap between the windows and circular weighting applied to each window. The results were de-noised using a 5 $\times$ 5 pixel median filter and any vectors with a correlation value of 0.1 were removed.}

\section{Results and Discussion}\label{sec:Results and Discussion} \label{sec:Results_and_Discussion}

We will now discuss the far-field sound and surface pressure fluctuations on the tip surface in tip clearance flows. \textcolor{black}{We will first consider the tip clearance flow-field and how it is influenced by the aerodynamic loading (angle of attack) and the clearance height. Next, the far-field sound results will be discussed followed by the surface pressure fluctuations on the tip.} In the case of the far-field sound, selected sound source maps will be shown first to establish some general characteristics of tip clearance noise observed under the present testing conditions. This will be followed by a discussion of the effect of Mach number, angle of attack and tip clearance height on the radiated sound through analysis of the integrated sound spectra. During the discussion of the far-field sound, we will utilise the nomenclature defined in \cref{fig:CBF_vs_DAMAS} (b) to highlight the behaviour of the different sound sources in the region of interest. Finally, the surface pressure statistics in the form of the root mean square ($RMS$) pressure, pressure spectra and two-point correlations at the same flow conditions as the far-field sound measurements will be presented. Unless otherwise stated, the results for both the far-field and near-field pressure fluctuations will be presented for $U_\infty$ = 30 m/s as those at other velocities are qualitatively the same. \par

\textcolor{black}{\subsection{Tip Clearance Flow-Field} \label{sec:PIV_Results}}

\begin{figure}
\centering
  \centerline{\includegraphics[width = \textwidth]{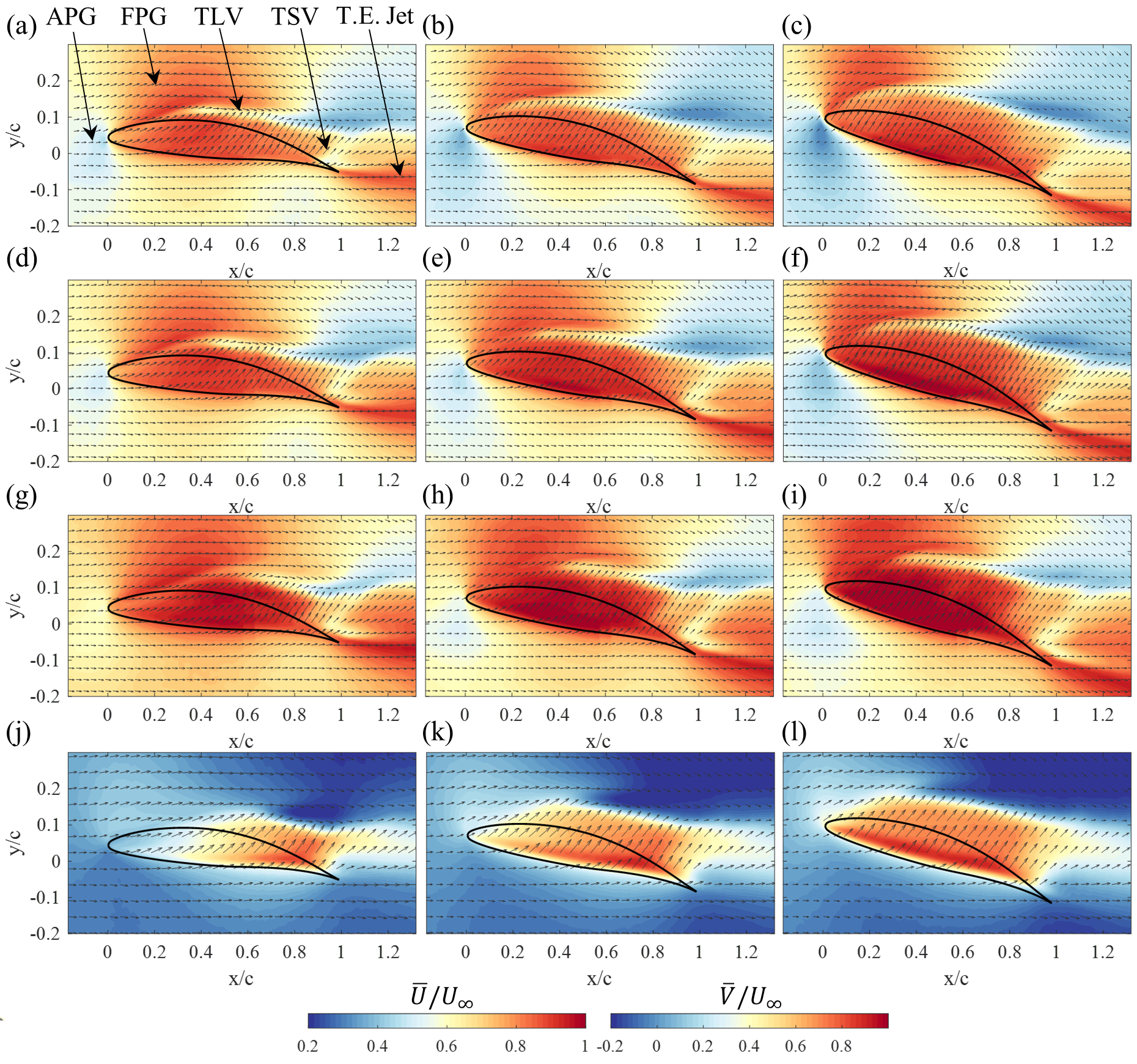}}
  \caption{\textcolor{black}{Streamwise (a -- i) and transverse (j -- l) mean velocity fields at $U_\infty$ = 15\,m/s in the wall-parallel plane from PIV measurements. The streamwise velocity distribution is shown for all three clearance heights: $h$ = 2.64\,mm (a,b,c); 3.96\,mm (d,e,f); 5.28\,mm (g,h,i) and all three angles of attack: $\alpha_g$ = 5.6$^\circ$ (a,d,g); 9.1$^\circ$ (b,e,h); 12.5$^\circ$ (c,f,i). The transverse velocity distribution is shown for $h$ = 5.28\,mm for $\alpha_g$ = (j) 5.6$^\circ$, 9.1$^\circ$ and 12.5$^\circ$. The projection of the airfoil and the mean streamlines have been superimposed upon each contour map with every 5$^{th}$ vector shown for clarity. The colourscale of each contour map is the same and shown at the bottom of the figure.}}
\label{fig:Mean_Velocity_Field}
\end{figure}

\textcolor{black}{\Cref{fig:Mean_Velocity_Field} shows the mean streamwise velocity field in the wall-parallel plane for each clearance height and angle of attack configuration considered in the present work. Some typical flow features present in each flow-field are shown in \cref{fig:Mean_Velocity_Field} (a). Near the leading-edge of the clearance, regions of adverse pressure gradient (APG) and favourable pressure gradient (FPG) develop towards the pressure and suction sides of the airfoil, respectively due to blockage. At a given angle of attack, the APG strength decreases as the clearance height increases (seen by comparing the maps along any column) and at a given clearance height it increases with angle of attack (seen by comparing the maps along any row). Similarly, the FPG strength increases as the gap height increases at a given angle of attack, and at a given clearance height, it also increases with aerodynamic loading. Further downstream, the leakage flow across the gap develops with its strength (revealed by the velocity magnitude and the streamlines) increasing as the angle of attack increases for a given clearance height and as the clearance height increases for a given angle of attack. The cross-flow rolls up into the leakage vortex (TLV) which is stronger and situated further away from the tip for larger angles of attack. The effect of the clearance height on the TLV is more subtle, but a reduction in the velocity deficit associated with the TLV as the clearance height increases can be clearly observed. Another region of velocity deficit, likely associated with the tip separation vortex (TSV), can be observed close to the trailing-edge. The TSV appears to be inherently linked to the TLV (as shown by the merging of the velocity deficit regions associated with the two features) which is reminiscent of the separation rib vortices revealed by the LES results of Koch \textit{et al.} \cite{RN253}. Finally, a jet-like feature can be observed exiting the gap near the trailing-edge for each case. }

\textcolor{black}{\Cref{fig:Mean_Velocity_Field} (j), (k) and (l) show the transverse velocity component in the gap for $h$ = 5.28\,mm at $\alpha_g$ = 5.6$^\circ$, 9.1$^\circ$ and 12.5$^\circ$, respectively. These plots and the streamline pattern show an increase in the cross-flow velocity through the gap as the angle of attack, and consequently the lift, increases. At the smallest angle of attack (\cref{fig:Mean_Velocity_Field} (j)), the strongest transverse velocity is located near the trailing-edge of the clearance, with weaker values in the mid-chord and the leading-edge regions. As the angle of attack increases further (\cref{fig:Mean_Velocity_Field} (k)), the velocity magnitude in the mid-chord region of the clearance is comparable to that in the trailing-edge region. Finally, at the largest angle of attack (\cref{fig:Mean_Velocity_Field} (l)), the lift is large enough such that the transverse velocity magnitude across the entire clearance (from the leading to the trailing-edge) is comparable. The increase in the cross-flow across the clearance in the mid-chord and the leading-edge regions has important implications in the far-field sound behaviour as discussed next.}

\begin{figure}
\centering
  \centerline{\includegraphics[width = \textwidth]{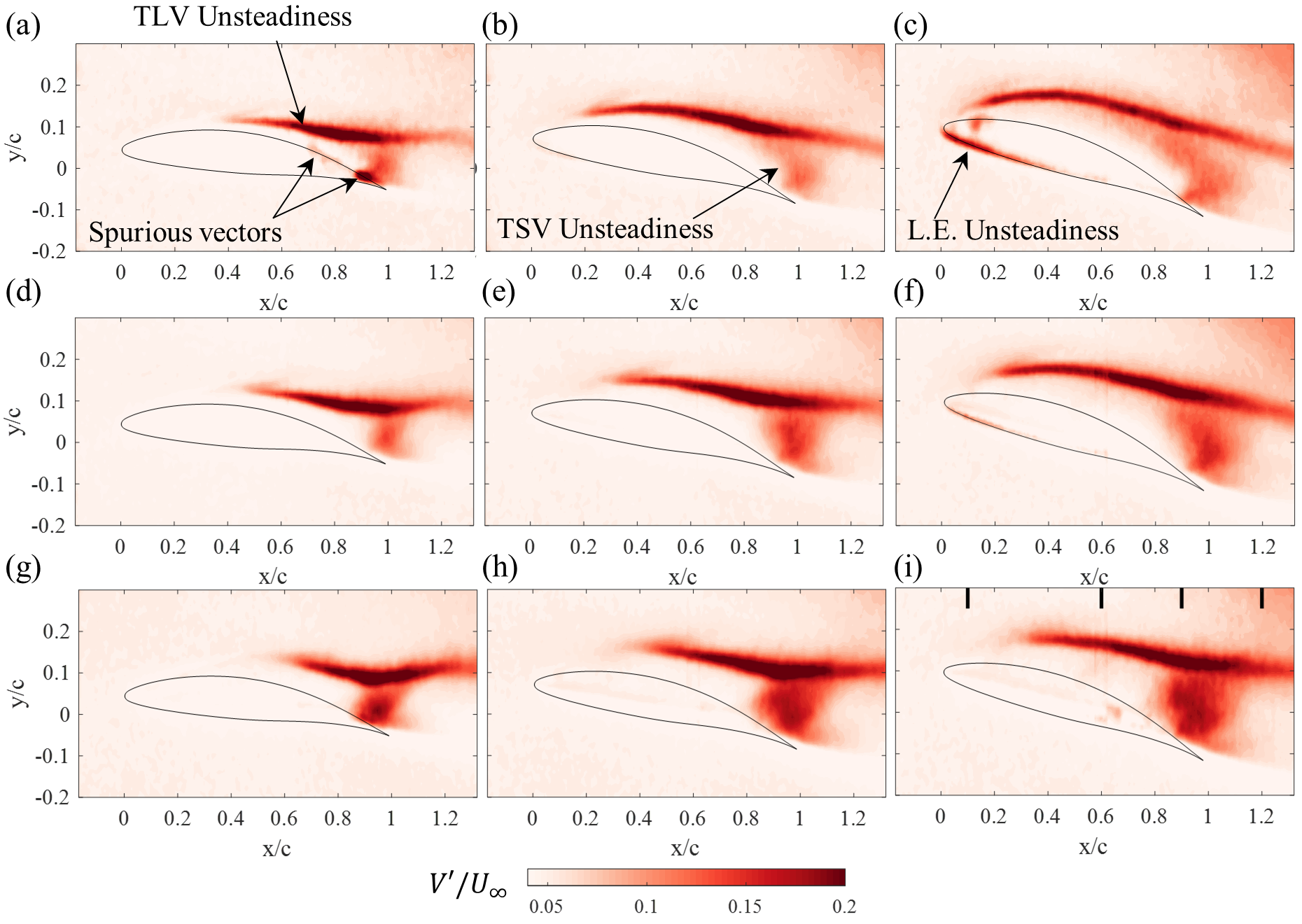}}
  \caption{\textcolor{black}{Root mean square (RMS) of transverse velocity at $U_\infty$ = 15\,m/s in the wall-parallel plane from PIV measurements. The RMS velocity distribution is shown for all three clearance heights: $h$ = 2.64\,mm (a,b,c); 3.96\,mm (d,e,f); 5.28\,mm (g,h,i) and all three angles of attack: $\alpha_g$ = 5.6$^\circ$ (a,d,g); 9.1$^\circ$ (b,e,h); 12.5$^\circ$ (c,f,i). The colourscale of each contour map is the same and shown at the bottom of the figure. The outline of the airfoil has been projected on each contour map for spatial reference. The four vertical bars near the top in (i) depict the streamwise locations for which RMS velocity profiles are shown in \cref{fig:RMS_Rake_Profiles}.}}
\label{fig:RMS_Velocity_Field}
\end{figure}

\textcolor{black}{\Cref{fig:RMS_Velocity_Field} shows the spatial distribution of the root mean square (RMS) of the transverse velocity component in the wall-parallel plane. These plots identify regions of large unsteadiness for all three clearance heights and angle of attack which is critical to understanding the far-field sound behaviour. Note, however, that because the PIV measurements are averaged over the finite thickness of the laser sheet (approximately 2\,mm) which is centred at the gap mid-height, these images may not paint a complete picture of the unsteady flow behaviour, particularly for larger clearance heights. Coming back to \cref{fig:RMS_Velocity_Field}, we first note that regardless of the lift or the clearance height the TLV and the TSV are the dominant source of unsteadiness in each case. For each clearance height, an increase in lift (observed by considering the maps along each row) pushes the TLV further away from the tip and larger velocity fluctuations are observed further upstream due to a stronger cross flow in the fore section of the clearance as shown by the mean velocity distribution above. For each angle of attack, the clearance height noticeably affects the TLV behaviour, as revealed by observing the maps down each column. The velocity fluctuations associated with the TLV increase with clearance height and an enhanced turbulence is observed closer to the tip surface for smaller clearance heights. The region of localised unsteadiness close to the trailing-edge associated with the TSV follows the TLV behaviour with velocity fluctuations increasing with an increase in an angle of attack and the clearance height. It is also interesting to note that, regardless of the flow configuration, the largest velocity fluctuations associated with the TSV are observed very close to the trailing-edge region. Lastly, we note that for the smallest clearance height at $\alpha_g$ = 12.5$^\circ$ (\cref{fig:RMS_Velocity_Field} (c)), a region of significant unsteadiness is also observed near the leading-edge along the edge of the tip toward the pressure-side. This is likely due to the oncoming boundary layer turbulence interacting with the gap, and as shown in the next section, results in a considerable mid-to-high frequency noise from this region. A similar, but weaker, leading-edge unsteadiness is also observed for the intermediate clearance height at $\alpha_g$ = 12.5$^\circ$ (\cref{fig:RMS_Velocity_Field} (f)). Finally, we note that although the velocity fluctuations inside the gap are generally lower than those within the vortex systems, it must be remembered that this could be an artefact of averaging over the laser sheet thickness and the fact that the sheet is centred at the gap mid-height. }

\begin{figure}
\centering
  \centerline{\includegraphics[width = \textwidth]{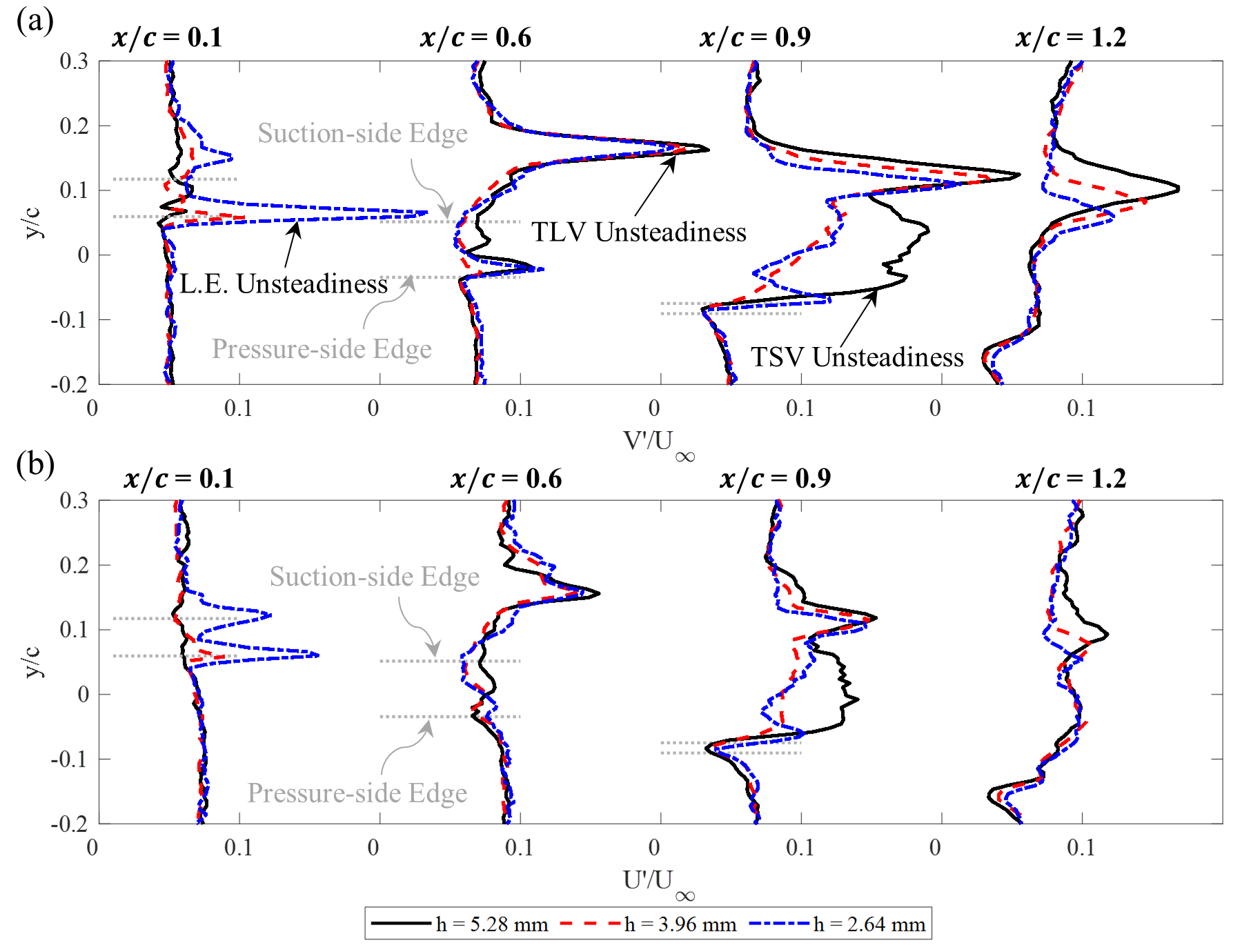}}
  \caption{\textcolor{black}{(a) Transverse RMS and (b) Streamwise RMS velocity profiles at $\alpha_g$ = 12.5$^\circ$ and $U_\infty$ = 15\,m/s for all three clearance heights along four vertical rakes placed at different streamwise stations (see \cref{fig:RMS_Velocity_Field} (i)). The streamwise location corresponding to each set of RMS velocity profiles is shown above the profile and the legend for each set of curves is shown at the bottom. The horizontal dashed lines on the profiles for $x/c \leq$ 0.9 depict the locations of the tip edges at the corresponding rake location.}}
\label{fig:RMS_Rake_Profiles}
\end{figure}

\textcolor{black}{To consider the quantitative effect of changing the clearance height on the velocity fluctuations across the clearance, \cref{fig:RMS_Rake_Profiles} (a) and (b) plots the RMS of the transverse and streamwise velocity components, respectively at four streamwise stations (highlighted in \cref{fig:RMS_Velocity_Field} (i)) for all three clearance heights at $\alpha_g$ = 12.5$^\circ$. Each set of curves here shows the RMS velocity profile for the three clearances at the streamwise location specified above each set. At each streamwise station, the edges of the tip in the vertical direction are represented using two dashed horizontal lines for each set of profiles. First consider the RMS velocity profiles of the transverse component shown in \cref{fig:RMS_Rake_Profiles} (a). The leading-edge unsteadiness observed in the contour map for $h$ = 2.64\,mm earlier manifests as peaks near the suction and pressure side edges at $x/c$ = 0.1, with the transverse velocity fluctuations just above the pressure-side edge being particularly intense. The magnitude of these peaks rapidly decreases with increasing clearance height, but a peak for 3.96\,mm can still be observed in the RMS value of both components. These peaks show that for small clearances, the interaction of the oncoming flow with the pressure-side edge results in generation of significant turbulence which, as shown later, is responsible for the generation of intense high-frequency noise radiation from this region.}

\textcolor{black}{Further downstream in the mid-chord region ($x/c$ = 0.6), the unsteadiness of the TLV dominates the RMS profiles with the magnitude and location of the peak being nearly independent of the clearance heights considered here. Inside the gap (the region bounded by the two horizontal lines), a peak near the pressure side edge of the tip due to flow separation there can be observed and slightly larger velocity fluctuations for the largest clearance ($h$ = 5.28\,mm) are observed closer to the suction-side edge of the tip. Closer to the trailing-edge ($x/c$ = 0.9), the peak in velocity fluctuations associated with the TLV begin to show effects of clearance height and an increase in clearance height increases the peak magnitude and shifts the peak location away from the tip. Besides the TLV-associated unsteadiness, this location also shows large broad peaks associated with the TSV just above the suction-side edge of the tip. As shown by the contour maps, the intensity of velocity fluctuations associated with the TSV decrease with the clearance height. It is shown later that this suppression of TSV-associated velocity fluctuations for smaller gap sizes results in a reduction of the low-frequency noise radiation from this region. Finally, at the farthest downstream station ($x/c$ = 1.2), the velocity fluctuations are dominated by the peaks associated with the TLV with the behaviour of these peaks being the same as that at $x/c$ = 0.9. Lastly, we note that the behaviour of the streamwise velocity fluctuations (\cref{fig:RMS_Rake_Profiles} (b)) at each measurement station is similar to that observed for the transverse component with the exception that their magnitude is noticeably weaker.} \par

\subsection{Tip Clearance Noise Characteristics: Analysis of Sound Source Maps} \label{sec:Source_Map_Analysis}

\begin{figure}
\centering
  \centerline{\includegraphics[width = \textwidth]{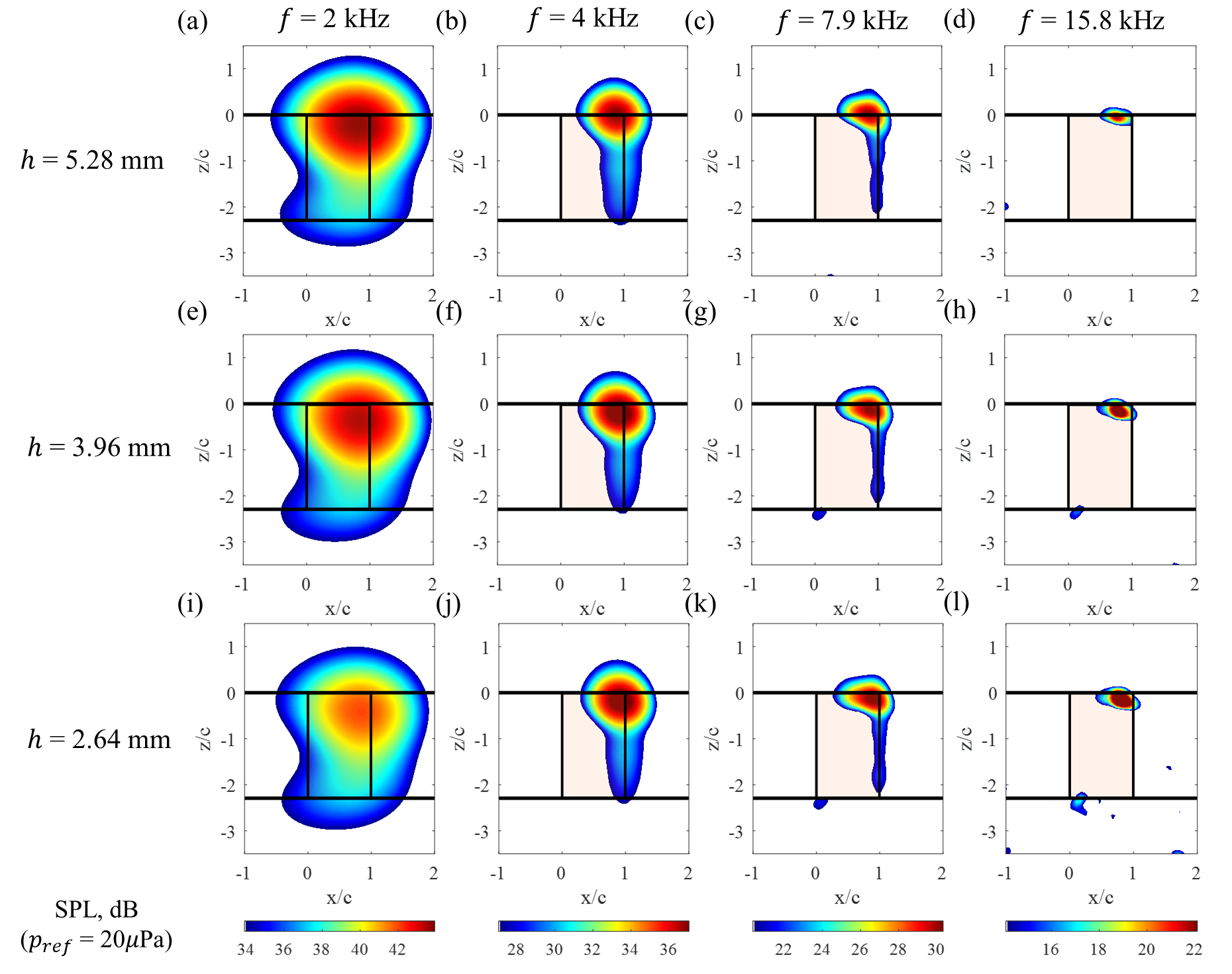}}
  \caption{The effect of changing the tip clearance height on the far-field sound sources at four frequencies for $\alpha_g$ = 5.6$^\circ$ and $U_\infty$ = 30\,m/s. Each row corresponds to a tip clearance height with (a) -- (d): $h$ = 5.28\,mm; (e) -- (h): $h$ = 3.96\,mm; (i -- l): $h$ = 2.64\,mm. Each column corresponds to the source maps at a particular frequency with (a, e, and i): 2 kHz; (b, f, and j): 4 kHz; (c, g, and k): 7.9 kHz; (d, h, and l): 15.8 kHz. The flow direction is from left-to-right in each map. The colour scale at each frequency is shown at the bottom of the corresponding column.}
\label{fig:DAMAS_Maps_Tip_Gap_Effect_5p6}
\end{figure}

\begin{figure}
\centering
  \centerline{\includegraphics[width = \textwidth]{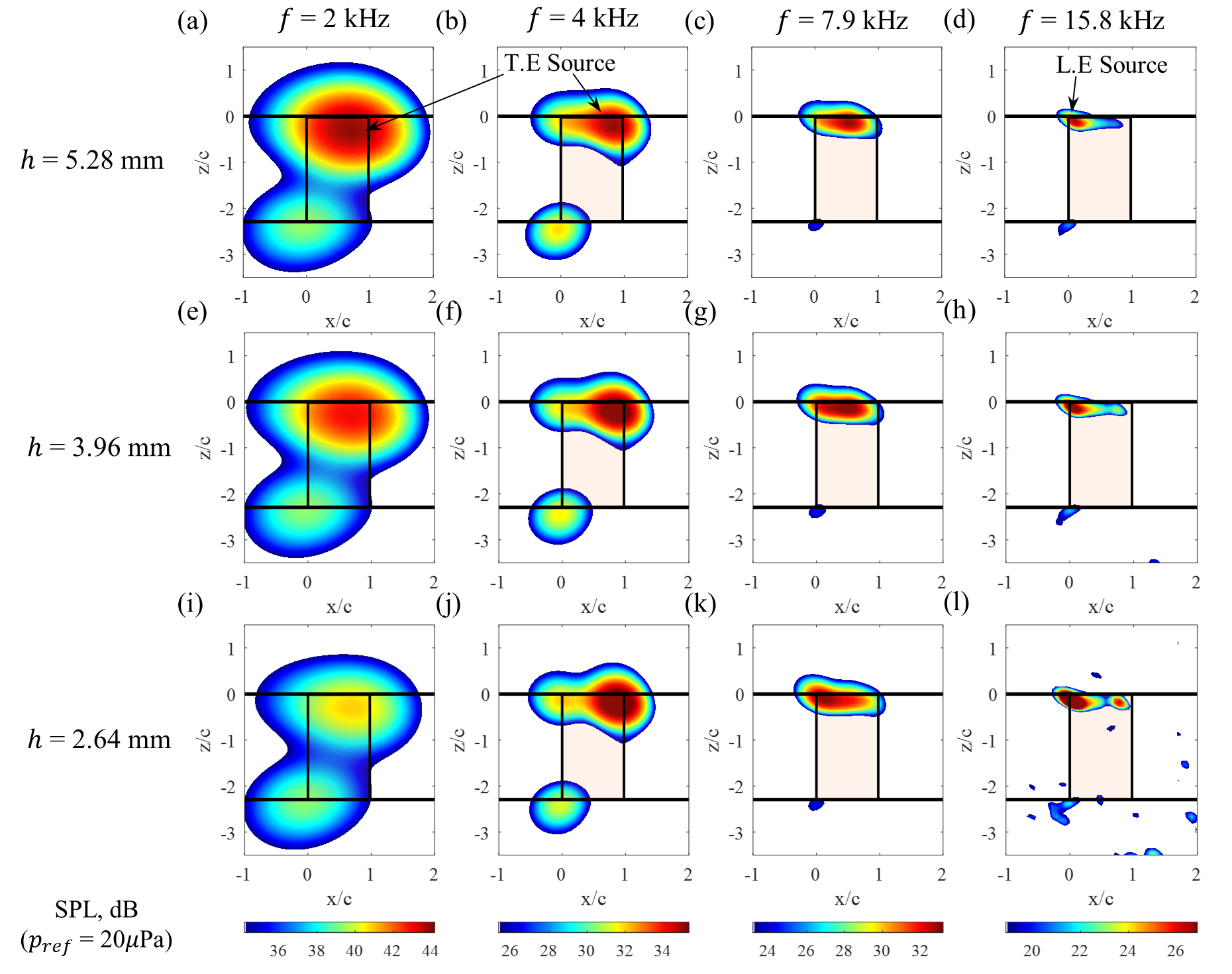}}
  \caption{The effect of changing the tip clearance height on the far-field sound sources at four frequencies for $\alpha_g$ = 12.5$^\circ$ and $U_\infty$ = 30\,m/s. The format of the figure is the same as that for \cref{fig:DAMAS_Maps_Tip_Gap_Effect_5p6}.}
\label{fig:DAMAS_Maps_Tip_Gap_Effect_12p5}
\end{figure}

We first consider some characteristics of the tip clearance noise through an analysis of the beamformed sound source maps. \textcolor{black}{We will utilise the sound-source maps obtained using the conventional beamforming algorithm as they are easier to visualise and interpret than the deconvolved DAMAS maps.} \Cref{fig:DAMAS_Maps_Tip_Gap_Effect_5p6} and \cref{fig:DAMAS_Maps_Tip_Gap_Effect_12p5} show a series of beamformed sound source maps for $\alpha_g$ = 5.6$^\circ$ and 12.5$^\circ$, respectively at four frequencies for the three tip clearance heights. The presentation of each source maps follows the same format as the source maps presented in \cref{fig:CBF_vs_DAMAS} in \cref{sec:Far Field Sound Instrumentation Description}. Each column in these figures shows the source maps at a particular frequency with frequency increasing from left-to-right. Each row corresponds to the maps for a particular tip clearance with the clearance height decreasing from top-to-bottom. At each frequency, the contour scale is locked between the maximum SPL and 15 dB down from the maximum SPL found in the map for the largest clearance ($h$ = 5.28 mm, first row). This allows both quantitative and qualitative comparison between the sound sources for the different tip clearance heights. \par

The effect of angle of attack on the clearance noise can be assessed by comparing the source maps for the largest tip clearance -- top rows in \cref{fig:DAMAS_Maps_Tip_Gap_Effect_5p6} and \cref{fig:DAMAS_Maps_Tip_Gap_Effect_12p5}. We first note that for both angles of attack the tip clearance noise is the dominant sound-source and it is clearly visible in each source map. Besides the clearance noise, the airfoil trailing-edge noise from the two-dimensional flow region in the mid-frequency range (4 kHz and 8 kHz) is clearly visible for the lowest angle of attack, but it is not prominent for $\alpha_g$ = 12.5$^\circ$ due to an increase in the relative magnitude of the clearance noise. Conversely, the airfoil leading-edge bottom wall junction noise (source near the bottom-left of the airfoil) is more prominent at $\alpha_g$ = 12.5$^\circ$ at 2\,kHz and 4\,kHz since the vortical junction flow becomes stronger as the angle of attack increases. \par

Now consider the behaviour of the tip clearance noise as a function of the angle of attack. At the lower angle of attack (\cref{fig:DAMAS_Maps_Tip_Gap_Effect_5p6} (a) -- (d)), the dominant clearance sound source at each frequency is located in the downstream half of the clearance ($x/c >$0.5). \textcolor{black}{ This is expected since, as shown by the PIV results in \cref{fig:Mean_Velocity_Field}, the strongest leakage flow under low angle of attack conditions resides in the downstream section of the tip.} Previous measurements of the flow-field around an idealised clearance flow formed by a NACA5510 blade at $\alpha_g$ = 5$^\circ$ \cite{RN189} have also shown that for such low angles of attack the highest turbulence levels are found in the downstream section. \textcolor{black}{As the angle of attack increases (\cref{fig:DAMAS_Maps_Tip_Gap_Effect_12p5} (a) -- (d)), the leading-edge and the mid-chord regions of the clearance become more efficient radiators of noise in the mid-to-high frequency range (7.9 kHz and 15.8 kHz). This behaviour is consistent with the PIV results which show that at $\alpha_g$ = 12.5$^\circ$ the leakage flow becomes stronger in the mid-chord and leading-edge regions with comparable transverse velocity magnitude as the trailing-edge region (see \cref{fig:Mean_Velocity_Field} (l)).} At lower frequencies (2 kHz and 4 kHz), sound sources in the leading-edge and the mid-chord regions of the tip can also be observed, but there is also a strong source near the trailing-edge (highlighted in \cref{fig:DAMAS_Maps_Tip_Gap_Effect_12p5} (a) and (b)) which is similar in appearance to that observed under lower angle of attack conditions. The effect of increasing the angle of attack on sound generated by smaller clearance heights appears similar, but as explained later in \cref{sec:Integrated_Spectra}, there are some key differences. \par

Now consider the effect of tip clearance height on the sound source at $\alpha_g$ = 5.6$^\circ$ (\cref{fig:DAMAS_Maps_Tip_Gap_Effect_5p6}). For this low angle of attack case, the loudest noise source remains closer to the trailing-edge of the tip clearance as the clearance height is reduced. At low frequencies (2 kHz, left column), a slight reduction in source strength is observed as the tip clearance is decreased, while in the mid-frequency range (4 kHz and 7.9 kHz) \textcolor{black}{a slight increase in noise levels with decreasing clearance height can be observed.} At high frequencies (15.8 kHz, last column), the largest source levels are also observed close to the trailing-edge of the tip clearance and a noticeable increase in the source levels with decreasing tip clearance height can be observed. As the angle of attack increases and the $TLF$ intensifies \textcolor{black}{as shown by the PIV results} (\cref{fig:DAMAS_Maps_Tip_Gap_Effect_12p5}), the sound source strength shows greater changes with tip clearance height. At the lowest frequency (2\,kHz, first column), the largest tip clearance sound source levels are observed near the \textcolor{black}{clearance trailing-edge which become noticeably weaker as the clearance height is reduced. This reduction in the source strength near the trailing-edge is likely a consequence of the reduction in TSV unsteadiness with clearance height as shown by the PIV results (see the RMS velocity profile at $x/c$ = 0.9 in \cref{fig:RMS_Rake_Profiles} (a)). For the higher frequencies (4\,kHz and 7.9\,kHz), the noise increases slightly with a reduction in the clearance height and a noticeable increase in the noise radiation from the leading-edge and the mid-chord regions is observed, particularly at 7.9\,kHz. Finally, at the highest frequency of 15.9\,kHz considered here, the most intense sound radiation occurs near the leading-edge and a reduction in the tip clearance height noticeably intensifies the sound source near the clearance leading-edge. This increase in noise from the leading-edge region for the smaller gaps is consistent with the PIV results which show elevated velocity fluctuations near the leading-edge for these configurations (see \cref{fig:RMS_Velocity_Field} (c) and (f)).} These source maps clearly show that the changes in the behaviour of clearance noise as the angle of attack and the clearance height changes are complex and frequency-dependent. To further understand the dependence of sound on clearance height, angle of attack and the frequency, we now turn to the far-field sound spectra obtained by integrating the sound source maps. \par

\subsection{Tip Clearance Sound Spectrum: The Effect of Mach Number, Angle of Attack and Clearance Height} \label{sec:Integrated_Spectra}

\textcolor{black}{We will now consider the Mach number ($M$) scaling of the tip clearance noise. However, before we do so, it must be emphasised that the purpose of presenting the scaling is not to establish a universal Mach number scaling. Rather, the results presented here serve to demonstrate the complex scaling behaviour of the tip clearance noise and its dependence on multiple parameters such as frequency, tip clearance height, angle of attack and even Reynolds number. We will consider the scaling of the total tip clearance noise that was obtained by integrating the source levels within the \textit{Tip clearance Region}.} \par

\textcolor{black}{Previous work on tip clearance noise has shown that its scaling behaviour is complex and a function of clearance height, however depending upon the frequency, two different mechanisms have been proposed \cite{RN189}. The first is a fifth power of velocity scaling at low frequencies that implies scattering of the turbulent fluctuations by the sharp edges of the tip and another is a $M^7 - M^8$ quadrupole scaling at higher frequencies that suggests a jet-like flow behaviour. In the present work we did not find the jet-like scaling to be applicable regardless of the clearance height or angle of attack configuration which is consistent with another recent work on aeroacoustics of an idealised tip clearance flow \cite{RN251}. Similar to this work, we believe that the absence of this scaling may be attributed to the lower Mach numbers in the present study. So we will instead focus on the $M^5$ scaling and clarify the conditions under which this scaling may be valid.} \par

\begin{figure}
\centering
  \centerline{\includegraphics[width = \textwidth]{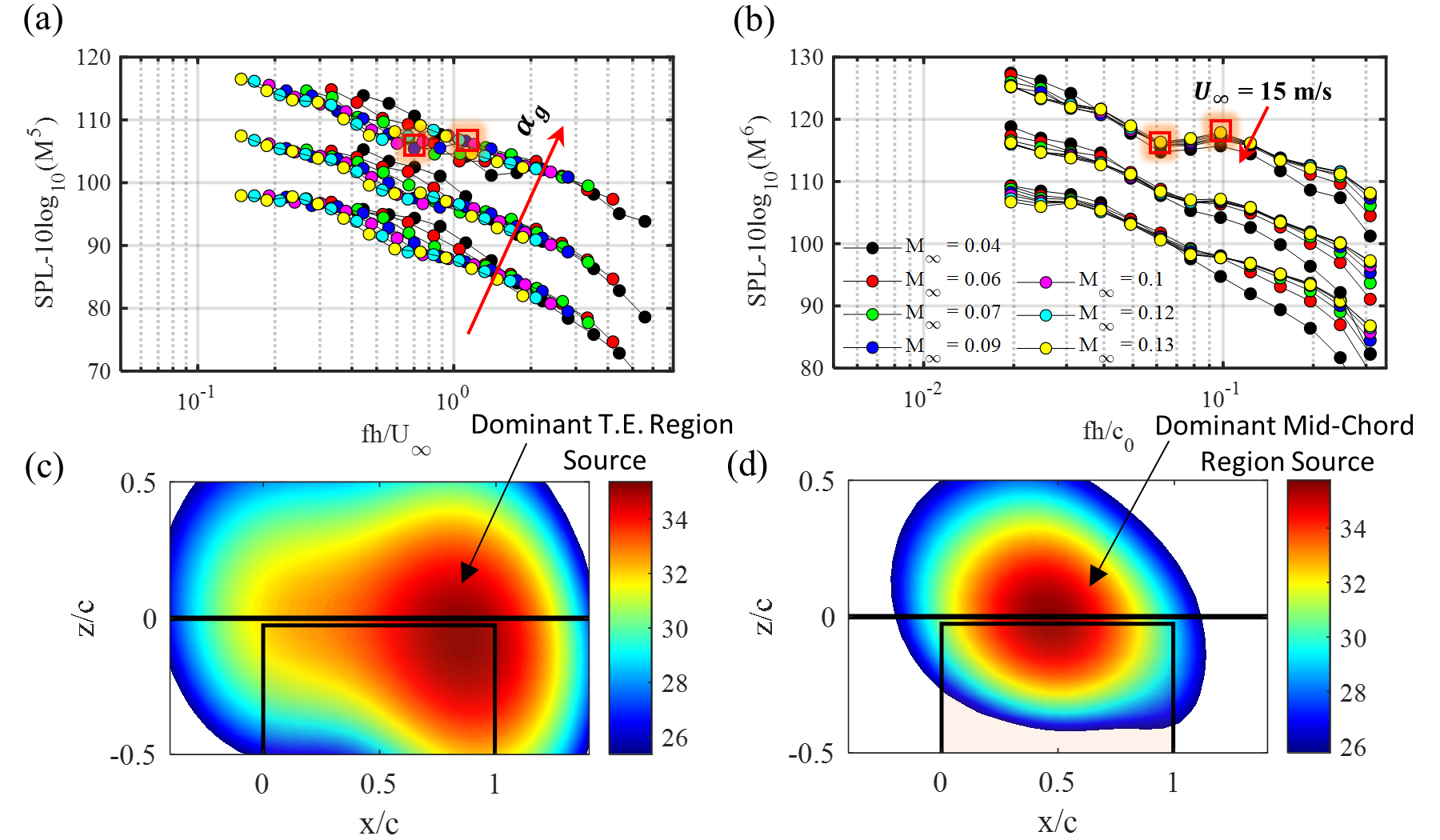}}
  \caption{\textcolor{black}{Mach number scaling of the total tip clearance noise for $h$ = 5.28\,mm and all three angles of attack. The tip clearance noise spectra scaled on the 5$^{th}$ power of Mach number and Strouhal number based on the free-stream velocity and clearance height is shown in (a), while (b) shows the scaling on $M^6$ and the Helmholtz number. The sound-source maps in (c) and (d) show the beamformed source map at 4\,kHz and 6.3\,kHz, respectively. The frequency corresponding to each source maps is also depicted using the rectangular boxes in (a) and (b). Note that an offset of 7.5\,dB and 15\,dB has been applied to each spectra for $\alpha_g$ = 9.1$^\circ$ and 12.5$^\circ$, respectively. Further note that the scale on y axis is fixed to 50\,dB for each plot presented here for consistency.}}
\label{fig:Velocity_Scaling_h5p3}
\end{figure}

\textcolor{black}{\Cref{fig:Velocity_Scaling_h5p3} (a) shows the total tip clearance noise for the largest clearance ($h$ = 5.28\,mm) with magnitude scaled on $M^5$ and frequency scaled on the clearance height and free-stream velocity for all three angles of attack. For each angle of attack configuration, this scaling collapses the spectra towards the lowest and the highest frequencies considered here, but a larger scatter (up to 5\,dB) that increases with the angle of attack, is observed in the mid-frequency range (approximately between $fh/U_\infty$ = 0.3 -- 1.5). A closer inspection of the individual spectra for the largest angle of attack (the set of curves on top) reveals that the larger scatter in the data in the mid-frequency range closely coincides with the mid-chord region of the clearance becoming a more efficient radiator of noise, compared to the trailing-edge. This shift in the dominant source region is clearly revealed by the beamforming maps in \cref{fig:Velocity_Scaling_h5p3} (c) and (d) that correspond to the two data points highlighted in \cref{fig:Velocity_Scaling_h5p3} (a). In fact, a similar behaviour, although less conspicuous, is also present in the far-field sound data for $\alpha_g$ = 9.1$^\circ$. For both angles of attack, the upstream shift in the dominant source region is associated with the formation of a spectral knee (which will become apparent momentarily) that is centred at about the same dimensional frequency (4\,kHz), regardless of the Mach number and thus scaling the frequency on the free-stream velocity leads to a larger scatter in the data. Considering the PIV results and the beamformed source maps discussed earlier, it is likely that an increase in the cross-flow intensity and noise levels in the mid-chord region for larger angles of attack lead to a change in the scaling behaviour of the tip clearance noise.} \par

\textcolor{black}{The presence of the spectral knee in the dataset whose frequency is independent of the flow velocity motivated another scaling of the tip clearance noise that is shown in \cref{fig:Velocity_Scaling_h5p3} (b). Here, the magnitude is scaled on $M^6$ (dipole scaling) and instead of the Strouhal number, the frequency is scaled on the clearance height and the ambient speed of sound (Helmholtz number). The motivation behind the use of Helmholtz number is the possibility that the formation of a velocity-independent spectral knee could be a result of significant acoustic scattering for the smaller clearances considered here. As shown in \cref{fig:Velocity_Scaling_h5p3} (b), this scaling collapses the data in the mid-frequency range (to within 1.2\,dB) for each angle of attack with the frequency range across which this scaling is valid increasing as both the angle of attack and the Mach number increase. For the lowest Mach number ($U_\infty$ = 15\,m/s), the scaling collapses the data only across a few frequency bands which could either be due to a Reynolds number effect or a low acoustic SNR for this case. The scaling presented here suggests that the tip clearance noise consists of a broad mid-frequency band where the magnitude scaling suggests a dipole behaviour, but the scaling of frequency on a constant time-scale (speed of sound in this case) suggests that the sound radiation is dominated by acoustic scattering, rather than aeroacoustic processes. This also implies that the frequency range over which this scaling is valid is non-compact in nature, despite the clearance height being acoustically compact. Note that the frequency range over which this scaling applies increases slightly with angle of attack and Reynolds number, presumably due to an increase in the intensity of acoustic scattering. Lastly, we note that although not shown, the spectra outside this mid-frequency range ($f<$ 3\,kHz and $f>$ 10\,kHz) do show a reasonable collapse when scaled on $M^4 - M^5$ and the Strouhal number based on the flow velocity, perhaps because aeroacoustic processes in these frequency ranges dominate the acoustic scattering.} \par

\textcolor{black}{Another noteworthy aspect of the spectra presented in \cref{fig:Velocity_Scaling_h5p3} (b) is that they clearly reveal the presence of the spectral knee mentioned earlier for the two larger angles of attack. Considering the beamforming source maps in \cref{fig:Velocity_Scaling_h5p3} (c) and (d) associated with the encircled data points shown in \cref{fig:Velocity_Scaling_h5p3} (b), it becomes clear that the formation of the spectral knee is associated with the mid-chord region (\cref{fig:Velocity_Scaling_h5p3} (d)) becoming a stronger sound radiator, compared to the trailing-edge region (\cref{fig:Velocity_Scaling_h5p3} (c)). Regardless of the Mach number, this occurs for frequencies greater than about 5\,kHz and 4\,kHz for $\alpha_g$ = 9.1$^\circ$ and 12.5$^\circ$, respectively. These results suggest that a stronger cross-flow in the mid-chord region (clearly visible in the transverse velocity distributions in \cref{fig:Mean_Velocity_Field} (k) and (l)) for larger angles of attack serves to increase the mid-frequency noise levels. This characteristic of the tip clearance noise will be discussed further during the discussion on the spectral levels from the three integration subregions defined in \cref{fig:CBF_vs_DAMAS} (b). Note that, although not shown, the frequency range over which $M^6$ scaling is applicable, it collapses the data regardless of whether the dominant source is located in the trailing-edge or the mid-chord region.} \par

%\begin{figure}
%\centering
%  \centerline{\includegraphics[width = \textwidth]{Velocity_Scaling_h5p3_Dipole}}
%  \caption{\textcolor{red}{Mach number scaling of the source levels in the (a) \textit{Mid-chord Region} and (b) \textit{T.E. Region} for $h$ = 5.28\,mm and $\alpha_g$ = 12.5$^\circ$. The sound-source maps in (e) and (f) correspond to the lower and higher frequencies, respectively encircled in (a) and (b).}}
%\label{fig:Velocity_Scaling_h5p3_Dipole}
%\end{figure}

\begin{figure}
\centering
  \centerline{\includegraphics[width = \textwidth]{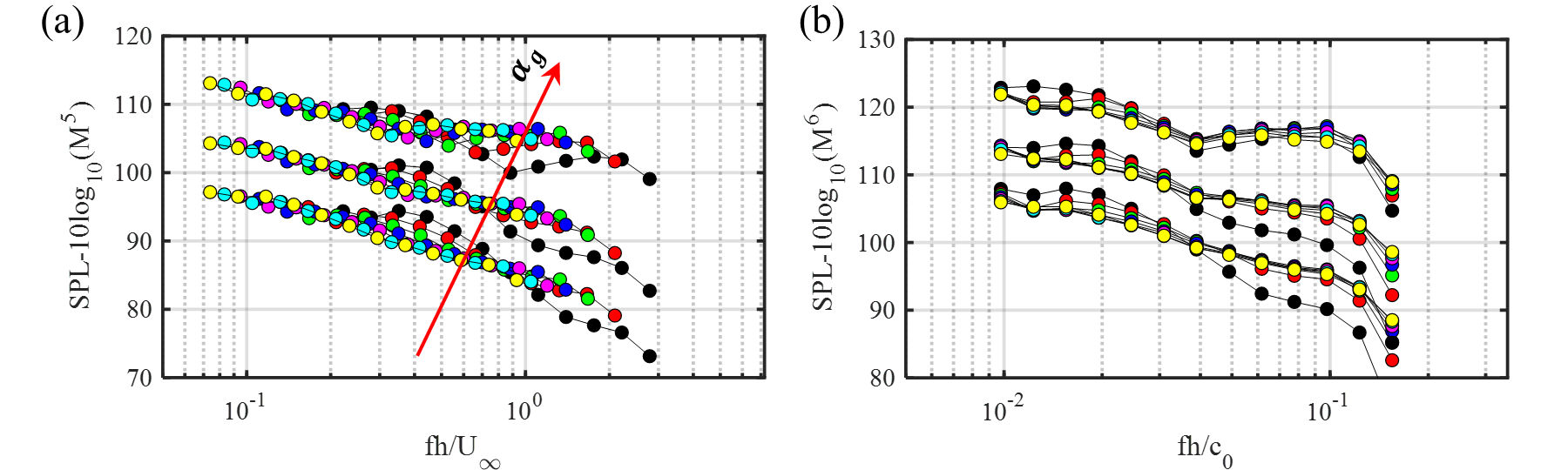}}
  \caption{\textcolor{black}{Mach number scaling of tip clearance noise for the smallest clearance height ($h$ = 2.64\,mm) at three different angles of attack. The scaling on $M^5$ and Strouhal number based on clearance height and free-stream velocity is shown in (a), while (b) shows the scaling on $M^6$ and the Helholtz number. The format of the plots is the same as \cref{fig:Velocity_Scaling_h5p3} (a) and (b).}}
\label{fig:Velocity_Scaling_Overall_Clearance_Noise_Smaller_Gaps}
\end{figure}

\textcolor{black}{Finally in our discussion on the Mach number scaling of the tip clearance noise, we consider the effect of lowering the clearance height on the scaling behaviour observed above for the largest clearance. \Cref{fig:Velocity_Scaling_Overall_Clearance_Noise_Smaller_Gaps} (a) and (b) show the $M^5$ and $M^6$ scaling, respectively applied to the overall tip clearance noise source levels for the smallest clearance ($h$ = 2.64\,mm). The scaling on $M^5$ for the smallest clearance shows a similar behaviour as the largest clearance (\cref{fig:Velocity_Scaling_h5p3} (a)) with the scaling only collapsing the data at lower and higher frequencies, and a larger scatter is observed in the mid-frequency range. The scaling on the $M^6$ and Helholtz number (\cref{fig:Velocity_Scaling_Overall_Clearance_Noise_Smaller_Gaps} (b)) is also similar as the largest clearance height with a good collapse of the levels observed across a broad mid-frequency range for each angle of attack. The scaling improves with Reynolds number and angle of attack as before and a larger scatter in the data is observed towards the lowest and highest frequencies. Overall, the scaling behaviour presented here shows that scaling the tip clearance noise spectra on Mach number alone may not be possible across the frequency range because of non-compactness effects that arise from scattering of the fluid dynamic noise source by the edges of the tip and the adjacent wall. It is also worth pointing out that the scaling on $M^6$ and Helmholtz number presented here may also be a function of clearance height since increasing the clearance height will also change the magnitude and frequency range over which strong acoustic scattering is observed.} \par

\begin{figure}
\centering
  \centerline{\includegraphics[width = \textwidth]{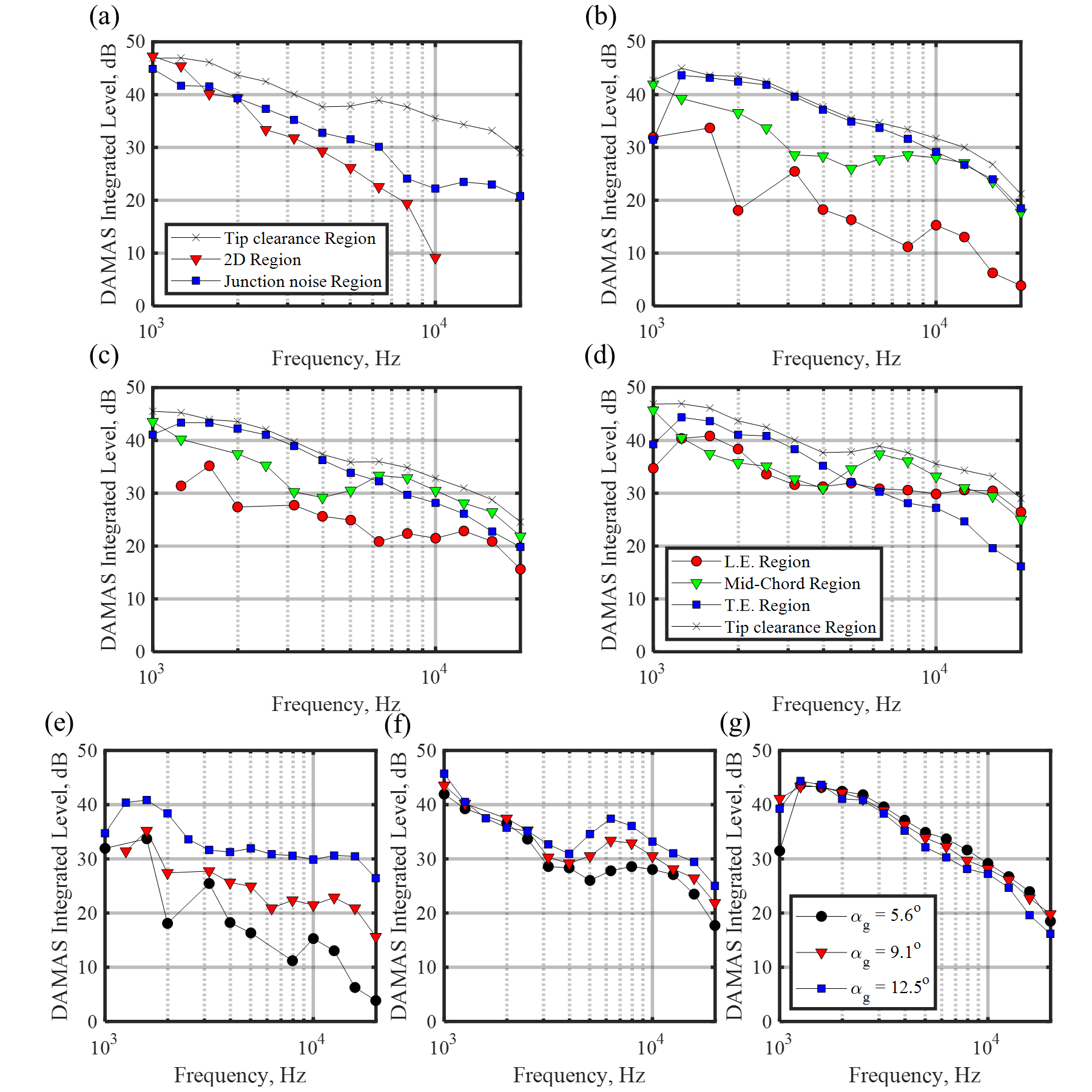}}
  \caption{The contribution of the clearance sub-regions specified in \cref{fig:CBF_vs_DAMAS} (b) to the far-field sound for $h$ = 5.28 mm at $U_\infty$ = 30 m/s. (a) The tip clearance noise relative to the airfoil trailing-edge noise and junction noise sources at $\alpha_g$ = 12.5$^\circ$. (b), (c), and (d) show the contribution of the three clearance sub-regions to the overall tip clearance noise for $\alpha_g$ = 5.6$^\circ$, 9.1$^\circ$, and 12.5$^\circ$, respectively. The legend for (b) -- (d) is shown in (d). The spectra for the \textit{L.E Region}, \textit{Mid-chord Region} and \textit{T.E. Region} as a function of angle of attack are shown in (e), (f) and (g), respectively. The legend for (e) -- (g) is shown in (g).}
\label{fig:DAMAS_Spectra_Region_Comparison}
\end{figure}

Now we consider the effect of changing the angle of attack on the noise levels within each of the clearance sub-regions with the $h$ = 5.28 mm clearance serving as our reference case. \textcolor{black}{In what follows, we will consider dimensional source spectra since we want to establish the effect of changing the lift (and later the clearance height) on the frequency content and magnitude of the far-field sound.} \Cref{fig:DAMAS_Spectra_Region_Comparison} (a) shows a comparison between the tip clearance noise and the two other major noise sources on the airfoil---the airfoil trailing-edge noise from the 2D flow region and the leading-edge bottom wall junction noise---for $\alpha_g$ = 12.5$^\circ$. As expected, the clearance noise exceeds the noise from other airfoil noise sources at most frequencies under the present flow conditions. The relative contribution of the three clearance sub-regions (defined in \cref{fig:CBF_vs_DAMAS} (b)) to the overall clearance noise for $\alpha_g$ = 5.6$^\circ$, 9.1$^\circ$, and 12.5$^\circ$ is shown in \cref{fig:DAMAS_Spectra_Region_Comparison} (b), (c), and (d), respectively. At the lowest angle of attack, the trailing-edge portion of the clearance ($x/c>$0.75) accounts for most of the clearance noise being generated. As the angle of attack increases to $\alpha_g$ = 9.1$^\circ$ (\cref{fig:DAMAS_Spectra_Region_Comparison} (c)), the trailing-edge portion still accounts for most of the radiated sound below about 6 kHz. However, at higher frequencies the noise contribution from the mid-chord region exceeds that from the trailing-edge region. At the highest angle of attack (\cref{fig:DAMAS_Spectra_Region_Comparison} (d)), a similar observation can be made with the exception that the mid-chord region source contribution exceeds the trailing-edge region contribution at frequencies greater than 4 kHz. Finally, for this high angle of attack case, for frequencies greater than 12 kHz the leading-edge region contribution to the noise becomes comparable to that from the mid-chord region. \textcolor{black}{These results, along with the PIV results and sound-source maps discussed earlier, suggest that a more intense cross-flow in the mid-chord and the leading-edge region as a result of a larger angle of attack significantly increases the mid-to-high frequency noise radiation, but the trailing-edge region continues to be a dominant source of low-frequency noise.} \par

The effect of the angle of attack on the clearance noise sources can be more clearly understood by plotting the spectra from the three sub-regions as a function of the angle of attack as shown in \cref{fig:DAMAS_Spectra_Region_Comparison} (e) -- (g). Considering these plots we can conclude that an increase in the strength of \textcolor{black}{the cross-flow in the leading-edge and mid-chord regions of the clearance with angle of attack (as shown by the PIV results) amplifies the mid-to-high frequency noise more as demonstrated by the strong variation of spectral levels with angle of attack observed in \cref{fig:DAMAS_Spectra_Region_Comparison} (e) and (f).} However, the low-frequency noise (between 1.2\,kHz and 4\,kHz) radiating from the trailing-edge region of the clearance is not as sensitive to the angle of attack as shown in \cref{fig:DAMAS_Spectra_Region_Comparison} (g). This is an important result since it leads to two conclusions. The first is that the turbulence generated by the $TLF$ in the mid-chord and leading-edge regions of the clearance controls the mid-to-high frequency noise, whereas the turbulence closer to the trailing-edge dictates the character of the low-frequency noise generation. The hypothesis that these are two distinct sound sources \textcolor{black}{is further supported by the fact the low-frequency noise shows a different scaling behaviour than the mid-to-high frequency clearance noise, see \cref{fig:Velocity_Scaling_h5p3} and associated discussion above.} The second conclusion is that the low-frequency, trailing-edge source strength is a weak function of the angle of attack, at least under the present conditions. This source is likely controlled by other parameters such as the clearance height or the tip geometry. In fact, as shown momentarily, the behaviour of the trailing-edge source is affected by the tip clearance height. Note that the trailing-edge source levels at 1\,kHz in \cref{fig:DAMAS_Spectra_Region_Comparison} (g) do show a strong angle of attack dependence, but as revealed by the deconvolved source maps, the source localisation at this frequency has large uncertainty in it due to the large array PSF for this frequency. \par
 
\begin{figure}
\centering
  \centerline{\includegraphics[width = \textwidth]{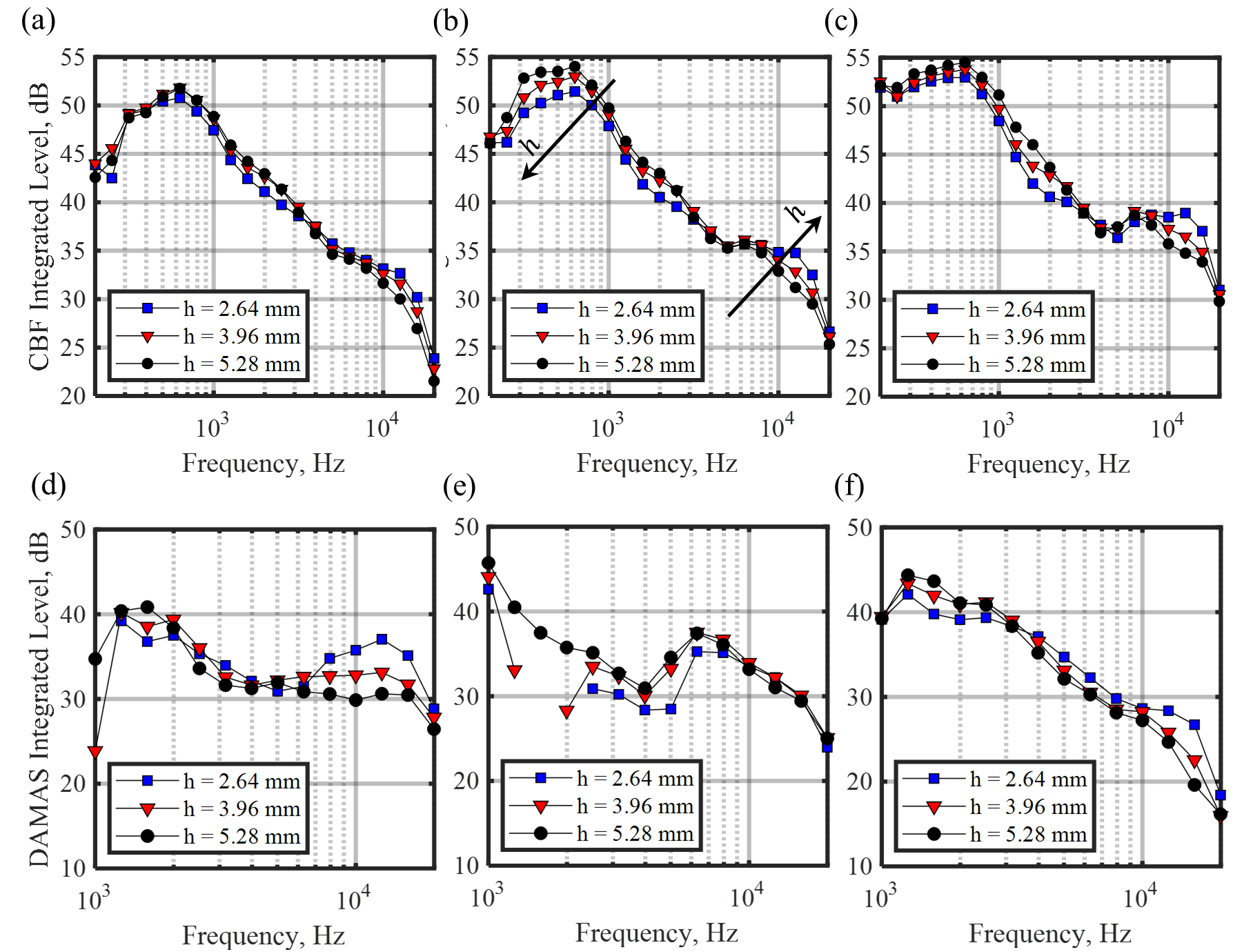}}
  \caption{The effect of tip clearance height on the far-field sound at $U_\infty$ = 30 m/s. (a), (b) and (c) show the total clearance noise as a function of the clearance height for $\alpha_g$ = 5.6$^\circ$, 9.1$^\circ$ and 12.5$^\circ$, respectively. (d), (e) and (f) show the effect of the tip clearance height on the sound radiated at $\alpha_g$ = 12.5$^\circ$ by the \textit{L.E. Region}, \textit{Mid-chord Region} and \textit{T.E. Region}, respectively. \textcolor{black}{Note that the spectra in (a) -- (c) that show the total clearance noise were obtained by integrating the CBF output, and therefore extend to lower frequencies.}}
\label{fig:DAMAS_Spectra_Tip_Gap_Comparison}
\end{figure}

We now present the effect of changing the tip clearance height on the far-field sound. \cref{fig:DAMAS_Spectra_Tip_Gap_Comparison} (a), (b) and (c) show the effect of changing the tip clearance height on the far-field sound at $U_\infty$ = 30 m/s for $\alpha_g$ = 5.6$^\circ$, 9.1$^\circ$ and 12.5$^\circ$, respectively. Note that these spectra were obtained by integrating the \textcolor{black}{beamforming, as opposed to DAMAS output,} over the entire tip clearance region (\textit{Tip clearance Region} in \cref{fig:CBF_vs_DAMAS} (b)). \textcolor{black}{The beamforming output is used here so that we can assess the effect the overall clearance noise down to the lowest frequencies (up to 200\,Hz) since the DAMAS output is less reliable below about 1.5\,kHz. Although the beamforming output also has large uncertainty at low frequencies with respect to its source localisation capabilities, it does allow us to assess the relative change in sound-source levels with clearance height even at low frequencies since the same array geometry and integration region definition are used to obtain the tip clearance spectra for each case.} Now consider the effect of changing the clearance height on the far-field sound spectra. At each angle of attack, the effect of reducing the tip clearance height appears to be qualitatively the same and three regimes can be identified. The first is a low-frequency regime (less than about 3 kHz) where a noticeable reduction in sound is observed with decreasing clearance height with the amount of noise reduction generally increasing as the angle of attack increases, \textcolor{black}{although for the highest angle of attack (\cref{fig:DAMAS_Spectra_Tip_Gap_Comparison} (c)) a slightly lower noise reduction (compared to $\alpha_g$ = 9.1$^\circ$) is observed below about 1\,kHz.} The second regime is a mid-frequency regime (approximately 3 -- 8 kHz) where the noise is less sensitive to the clearance height. Finally, there exists a high-frequency regime (10\,kHz and above) where the noise increases with decreasing tip clearance, with the amount of noise amplification increasing as the angle of attack increases. This increase in high-frequency noise for smaller clearances has been observed previously in measurements of clearance noise from a ducted fan \cite{RN270}. Additionally, the fact that the effect of tip clearance on the noise is frequency-dependent has also been demonstrated in recent measurements by Palleja-Cabre \textit{et al.} \cite{RN251}. \par

We will now show that the effect of changing the tip clearance height on the radiated sound is a function of frequency. As noted previously, different regions of the clearance control different parts of the sound spectrum and changing the clearance height modifies the sound source in these regions differently. For the purpose of this analysis we will only consider the sound generated at the highest angle of attack for $U_\infty$ = 30 m/s and \textcolor{black}{we revert back to the DAMAS results due to their better spatial resolution that help localise the sound sources in the three sub-regions of the tip clearance.} \Cref{fig:DAMAS_Spectra_Tip_Gap_Comparison} (d), (e) and (f) plot the spectra from the leading-edge, the mid-chord and the trailing-edge sub-regions, respectively as a function of clearance height for $\alpha_g$ = 12.5$^\circ$. We first note that the shape of the spectra and the effect of clearance height on the far-field sound is different in the three sub-regions considered here. In the leading-edge region (\cref{fig:DAMAS_Spectra_Tip_Gap_Comparison} (d)), reducing the clearance height results in an amplification of high-frequency noise greater than about 7\,kHz. Considering the sound source maps shown in the previous section (last column in \cref{fig:DAMAS_Maps_Tip_Gap_Effect_12p5}), we can see that this amplification is a result of the source region becoming concentrated close to the leading-edge for smaller clearances. This increase in high frequency noise from the leading-edge region is responsible for the sharp increase in the overall high frequency clearance noise with decreasing clearance height observed in \cref{fig:DAMAS_Spectra_Tip_Gap_Comparison} (c). \textcolor{black}{Considering the RMS transverse velocity field for the three clearance heights at $\alpha_g$ = 12.5$^\circ$ (see \cref{fig:RMS_Velocity_Field} (c), (f) and (i)), we can hypothesise that the high-frequency noise from the smaller clearances is a result of the oncoming turbulence violently interacting with the edges of the tip.} \par

In contrast to the sound radiation from the leading-edge region, the high frequency noise above about 7\,kHz from the mid-chord region (\cref{fig:DAMAS_Spectra_Tip_Gap_Comparison} (e)) remains nearly independent of the clearance height, while the sound at lower frequencies in this region decreases as the clearance height is reduced. \textcolor{black}{This is, to some extent, consistent with the PIV results which shows larger streamwise and transverse velocity fluctuations for the largest clearance in the mid-chord region (see the RMS velocity profiles at $x/c$ = 0.6 in \cref{fig:RMS_Rake_Profiles}). It is also worth noting that smaller clearances have been shown to shift the strongest $TLF$ further upstream towards the leading-edge in axial compressor flow \cite{RN261}, an effect that is also visible in the present PIV results for the largest angle of attack configuration (see the RMS velocity profiles in \cref{fig:RMS_Rake_Profiles}, for example). It is therefore likely that the amplification of the high frequency sound in the leading-edge region and a lowering of the low-to-mid frequency sound from the mid-chord region for the smallest clearance is related to this upstream shift in the $TLF$ intensity.} \par

Finally, in the trailing-edge region (\cref{fig:DAMAS_Spectra_Tip_Gap_Comparison} (f)) the sound up to approximately 3\,kHz decreases with clearance size, while an increase in high-frequency noise, similar to that observed near the leading-edge but lower in overall magnitude, is also observed here. The sound in the mid-frequency range (4 -- 8 kHz) is a weaker function of clearance height, but a slight increase in the source levels with a reduction in clearance height can be observed. Also note that the low-frequency noise ($<$ 3\,kHz) from the trailing-edge region is generally larger in magnitude than the sound from the regions further upstream and the spectral behaviour of the sound in this region is similar to that of the overall clearance noise behaviour shown in \cref{fig:DAMAS_Spectra_Tip_Gap_Comparison} (c). This is not surprising given that it has already been shown that, regardless of an increase in the angle of attack, the trailing-edge portion of the clearance dominates the noise production at low frequencies (see \cref{fig:DAMAS_Spectra_Region_Comparison} (b) -- (d)). The fact that this low-frequency noise source is a stronger function of clearance height than the angle of attack implies that the former could be a more important design parameter to consider when low-frequency noise reductions are desired. It is also worth noting that the presence of three separate sound source mechanisms suggested here may also be a function of the airfoil shape. However, the possibility of the existence of three separate source mechanisms in clearance flows has also been suggested recently by Palleja-Cabre \textit{et al.} \cite{RN251} who measured the far-field sound from an idealised tip clearance with a NACA 6512-10 airfoil. This suggests that the present results may be, at least qualitatively, applicable to situations where the tip clearance is formed by placing a stationary cambered airfoil adjacent to a flat plate. \par

\begin{figure}
\centering
  \centerline{\includegraphics[width = \textwidth]{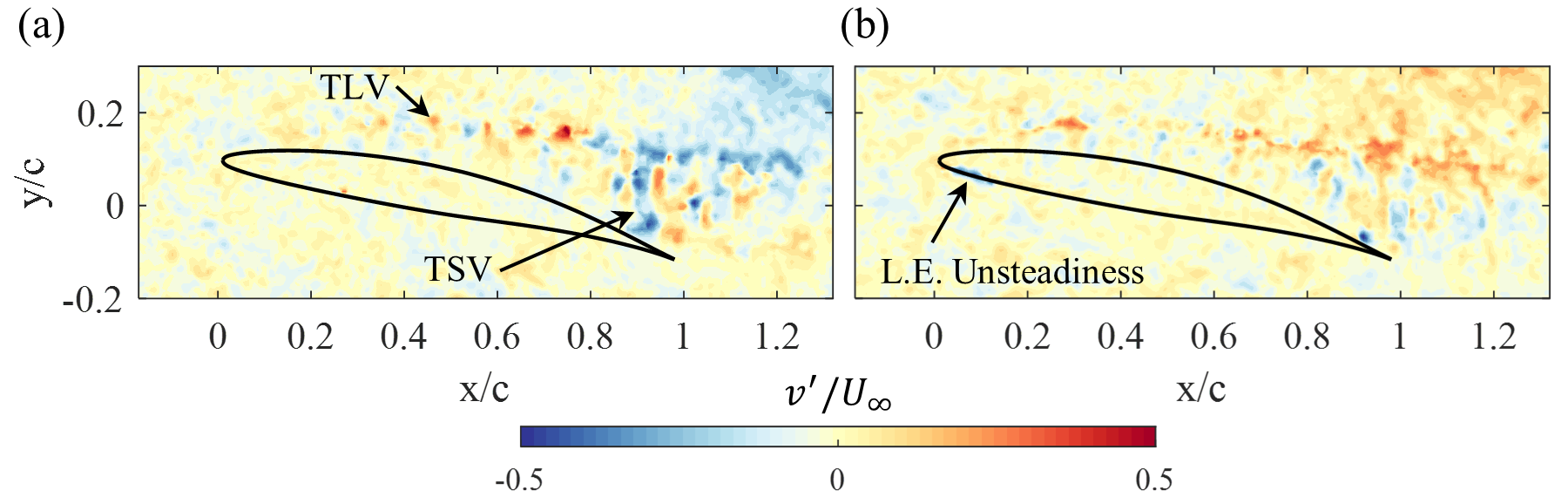}}
  \caption{\textcolor{black}{Tip separation vortices for $h$ = (a) 5.28\,mm and (b) 2.64\,mm at $\alpha_g$ = 12.5$^\circ$ visualised through instantaneous, mean-subtracted PIV snapshots of the transverse velocity component The colourbar for each plot is the same and shown at the bottom.}}
\label{fig:Instantaneous_Velocity}
\end{figure}

Before we examine the behaviour of the near-field pressure in clearance flows, a few statements regarding the possible origin of the low-frequency, trailing-edge source in clearance flows are warranted. It is apparent that this low-frequency source is not directly related to the \textcolor{black}{cross-flow through the clearance in the mid-chord and leading-edge regions} since it is only weakly affected by the angle of attack \textcolor{black}{which increases the cross-flow intensity in these regions}. On the other hand, the fact that the clearance height has a stronger influence on the strength of this source suggests that it could be related to the tip separation vortex ($TSV$) which is formed as the flow in the gap separates under the influence of the wall. \textcolor{black}{In fact, the RMS transverse velocity contour maps from PIV shown earlier in \cref{fig:RMS_Velocity_Field} show that the magnitude of the largest velocity fluctuations in the trailing-edge region are a much stronger function of clearance height (see maps along the columns) than the angle of attack (see maps along the rows). These velocity fluctuations that are associated with the $TSV$ presumably dominate the low-frequency production of noise from this region and the reduction in their magnitude with clearance height results in a reduction of the low-frequency noise radiation from the trailing-edge region. To further illustrate this point, \cref{fig:Instantaneous_Velocity} (a) and (b) show the instantaneous transverse velocity fluctuation snapshots at $\alpha_g$ = 12.5$^\circ$ for $h$ = 5.28\,mm and 2.64\,mm, respectively. The coherent structures associated with the $TLV$ and $TSV$ are clearly visible in these snapshots, along with the unsteadiness near the leading-edge for the smaller clearance. The large coherent structures associated with the $TSV$ visible here can be expected to generate significant low-frequency noise due to their proximity to the sharp edge of the tip which explains why the trailing-edge region remains the dominant low-frequency noise source (see source maps at 2\,kHz and 4\,kHz in \cref{fig:DAMAS_Maps_Tip_Gap_Effect_12p5}) despite an increase in the intensity of the bulk cross-flow through the gap further upstream as the angle of attack increases. Comparing \cref{fig:Instantaneous_Velocity} (a) and (b), we note a clear reduction in the strength of the coherent structures associated with both vortex systems, which explains the reduction in the low frequency noise from the trailing-edge region as the clearance height is reduced.} 

\subsection{Analysis of Tip Clearance Sound Source Region: Surface Pressure Fluctuations on the Tip} \label{sec:Surface_Pressure_Fluctuations}

We now consider the surface pressure fluctuations measured on the tip surface along the camber line of the airfoil at 15 different locations previously listed in \cref{tab:Surf_Pressure_Locations}. The frequency range of present measurements is between 200 Hz and 10 kHz since the measurements below 200 Hz are contaminated by the facility noise, while those above 10 kHz suffer from low signal-to-noise ratio in both measurements and calibrations. We will first consider the effect of changing the angle of attack on the pressure statistics for the largest tip clearance before discussing the effect of changing the clearance height. \par

\begin{figure}
\centering
  \centerline{\includegraphics[width = \textwidth]{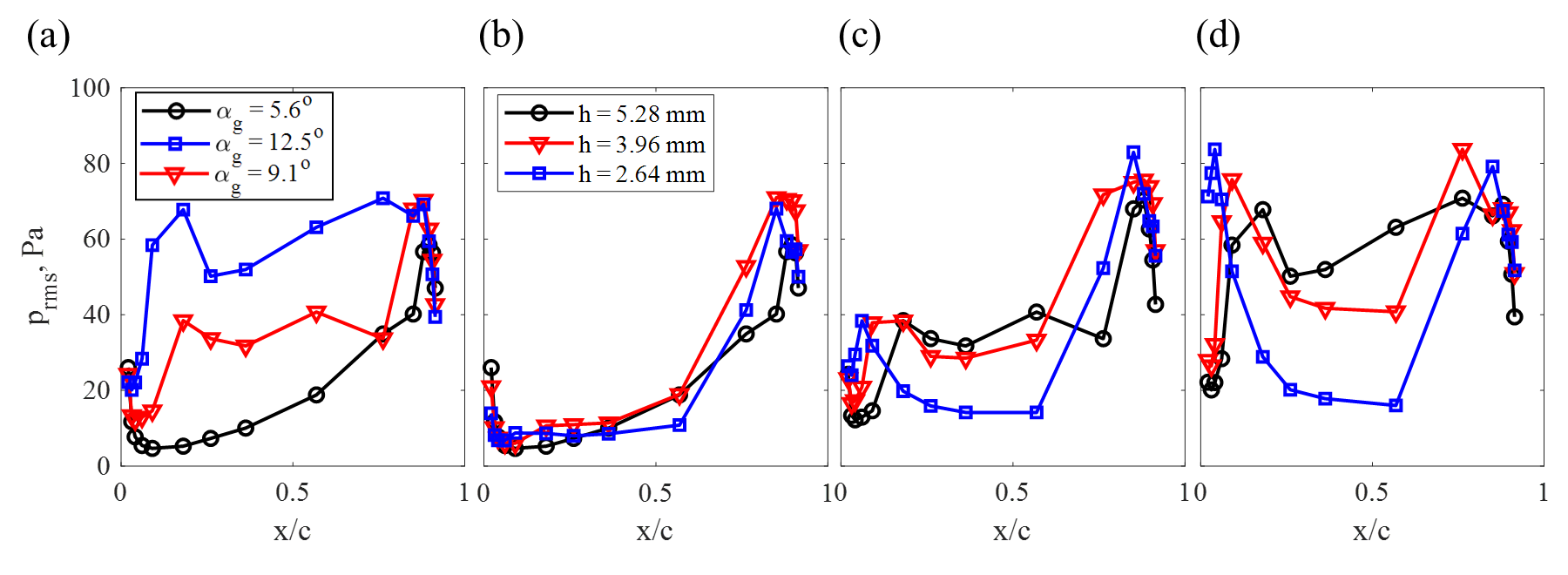}}
  \caption{The root mean square ($RMS$) pressure at $U_\infty$ = 30 m/s on the tip surface as a function of (a) angle of attack for $h$ = 5.28 mm and as a function of tip clearance for $\alpha_g$ = (b) 5.6$^\circ$, (c) 9.1$^\circ$ and (d) 12.5$^\circ$. The legend for (b) -- (d) is shown in (b).}
\label{fig:Tip_PRMS}
\end{figure}

\Cref{fig:Tip_PRMS} (a) shows the $RMS$ pressure obtained by integrating the pressure spectra between 200 Hz and 10 kHz for $h$ = 5.28 mm for the three angles of attack as a function of the chordwise measurement location. For the smallest angle of attack, the pressure decays slightly in the chordwise direction near the leading edge before rising nearly exponentially to a peak around $x/c$ = 0.9 after which a sharp drop-off is observed in proximity to the trailing-edge. Increasing the angle of attack has little influence on fluctuating pressure near the trailing-edge, however, it leads to a large amplification of pressure fluctuations in the leading-edge and the mid-chord region due to the $TLF$ becoming stronger as the lift increases. This trend is exactly the same as the trend in the far-field sound where the sound radiation from the trailing-edge region was found to be nearly independent of the angle of attack, but an increase in the angle of attack amplified the mid-to-high frequency noise radiating from the leading-edge and the mid-chord regions. \par

\Cref{fig:Tip_PRMS} (b) -- (d) show the effect of tip clearance height on the $RMS$ pressure along the chord-wise direction for $\alpha_g$ = 5.6$^\circ$, 9.1$^\circ$ and 12.5$^\circ$, respectively. At the lowest angle of attack (\cref{fig:Tip_PRMS} (b)), the fluctuating surface pressure is more sensitive to the clearance height in the downstream half of the clearance, particularly for $x/c>$ 0.75 where a rise in pressure levels is observed with decreasing clearance height. In close proximity to the leading-edge ($x/c$ = 0.02), a slight lowering of pressure levels with decreasing clearance height is also observed. As the angle of attack increases (\cref{fig:Tip_PRMS} (c) and (d)), the fluctuating pressure becomes a strong function of the clearance height. Near the leading-edge, a reduction in clearance height moves the local maximum in this region further upstream and also increases the value of this maximum slightly in case of $\alpha_g$ = 12.5$^\circ$ (\cref{fig:Tip_PRMS} (d)). \textcolor{black}{This is expected since the PIV results show that a lowering of clearance height results in a more intense $TLF$ further upstream.} In the mid-chord region, decreasing the clearance height results in a substantial reduction of fluctuating pressure for $0.1<x/c<0.75$. Closer to the trailing-edge the fluctuating pressure for lower clearances shows a sharper peak with a slightly higher maximum value. Overall, it appears that for large enough angle of attack, reducing the clearance height concentrates the most intense pressure fluctuations in close proximity to the leading and trailing-edges of the clearance, and it also results in a substantial reduction of fluctuating pressure in the mid-chord region. \par  

\begin{figure}
\centering
  \centerline{\includegraphics[width = \textwidth]{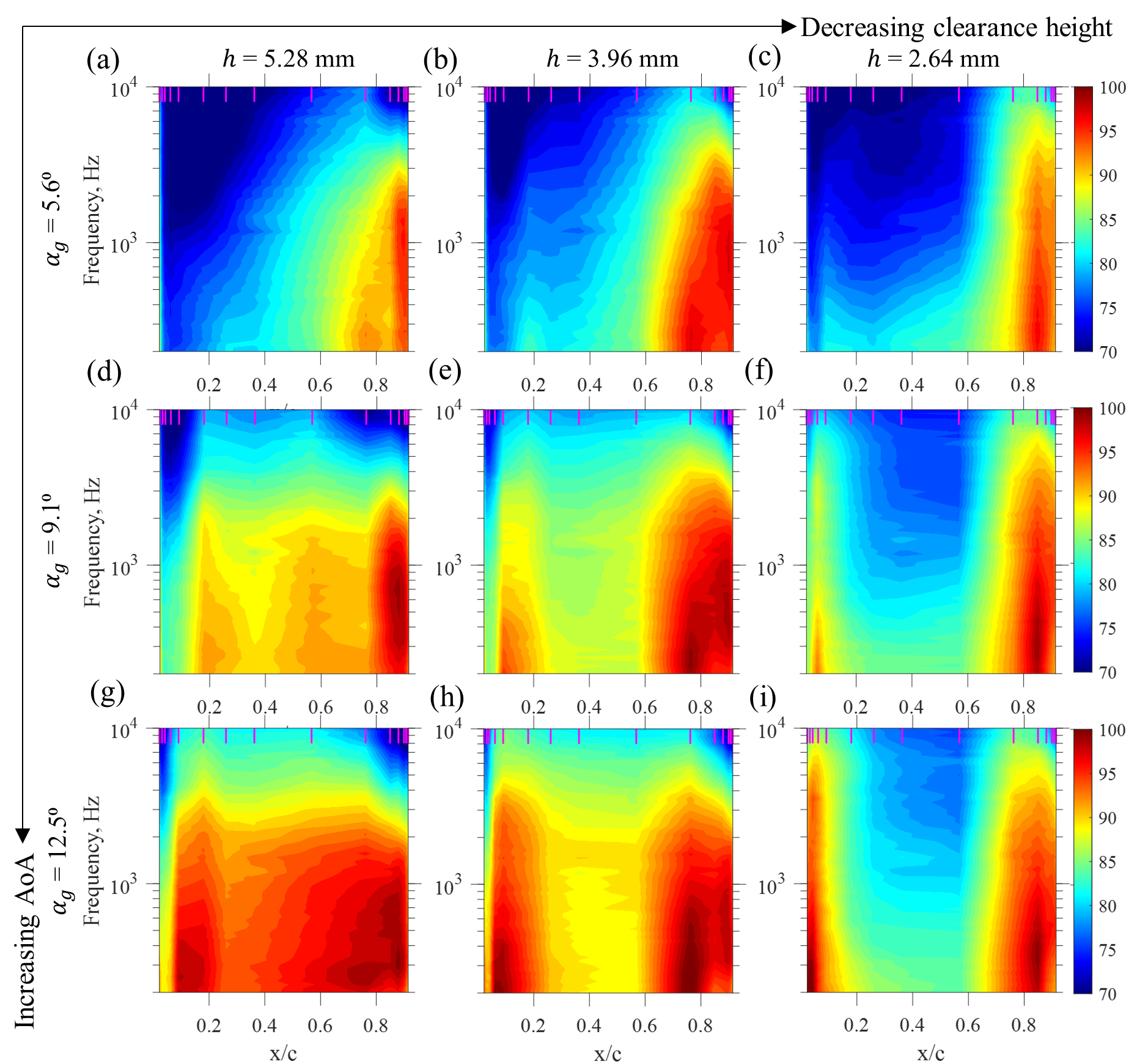}}
  \caption{Contour maps of spectral density of surface pressure fluctuations (in dB/Hz) on the tip surface for three angles of attack and three tip clearance heights at $U_\infty$ = 30 m/s. The contour maps in the first, second and third column correspond to $h$ = 5.28 mm, 3.96 mm and 2.64 mm, respectively, while the first, second and third row correspond to maps for $\alpha_g$ = 5.6$^\circ$, 9.1$^\circ$ and 12.5$^\circ$, respectively. \textcolor{black}{The vertical markers at the top within each map represent the chordwise location of the discrete remote microphone sensors.}}
\label{fig:Tip_Spectral_Contours}
\end{figure}

\Cref{fig:Tip_Spectral_Contours} shows the fluctuating pressure (in dB/Hz) on the tip surface as a function of both the chordwise distance (horizontal axis) and the frequency (vertical axis) for all three angles of attack and tip clearances. The colour scale (shown to the right) of these maps has been fixed so that the effects of both the angle of attack and the clearance height on the surface pressure fluctuations can be assessed qualitatively and quantitatively. The effect of angle of attack for a particular clearance height can be analysed by comparing the contour maps along any column, while the effect of the clearance height for a particular angle of attack configuration can be assessed by comparing the maps along any row. \textcolor{black}{Note that because these maps were created by interpolating the spectral data between the discrete measurement locations (represented by the short vertical bars at the top of each map), these plots should not be interpreted as being representative of the unsteady surface pressure behaviour across the entire tip surface. Instead, the aim of these spectral maps is to illustrate the effect of changing the lift and the clearance height on the pressure fluctuations in different regions of the tip clearance and establish their correlation with the behaviour of the sound-source in these regions discussed earlier.} Coming back to \cref{fig:Tip_Spectral_Contours}, we note that at the lowest angle of attack (first row) the highest pressure fluctuations are located near the trailing-edge of the clearance which is why this region accounts for most of the noise generated for this case. A reduction in clearance size for this case does not change the peak value of pressure fluctuations significantly, but it concentrates the most intense fluctuations (observed at low frequencies) close to the trailing-edge and also increases the magnitude of high-frequency pressure fluctuations. A slight increase in pressure fluctuations in the low-to-mid frequency range with lowering of clearance height can also be observed near the leading-edge. However, it is the increase in high-frequency pressure fluctuations near the trailing-edge which is responsible for an increase in high-frequency sound observed in \cref{fig:DAMAS_Spectra_Tip_Gap_Comparison} (a) for this low angle of attack case. \par

As the angle of attack increases (second and third rows), a broadband increase in pressure fluctuations in the mid-chord and the leading-edge portions of the clearance can be clearly observed, however, the trailing-edge portion of the clearance remains an important source of low-frequency pressure fluctuations. This explains the dominance of this region in the production of low-frequency noise, regardless of the angle of attack. The most obvious effect of lowering the clearance height at high angle of attack is that it concentrates the regions of highest pressure fluctuations towards the leading and trailing-edges of the clearance, while lowering the pressure fluctuations in the mid-chord region. At lower frequencies (below about 3 kHz), where the trailing-edge portion is the dominant far-field sound source, the reduction in the area over which the highest pressure fluctuations are observed as the clearance is reduced is likely responsible for the reduction in the far-field sound levels observed in the previous section (\cref{fig:DAMAS_Spectra_Tip_Gap_Comparison} (f)). \textcolor{black}{Considering the PIV results shown earlier, this reduction in both near and far-field pressure fluctuations with clearance height is linked to a reduction in the turbulence associated with the $TSV$ in the trailing-edge region.} Similarly, the reduction in far-field sound with decreasing clearance height below about 7 kHz observed earlier (\cref{fig:DAMAS_Spectra_Tip_Gap_Comparison} (e)) is related to the weakening of the pressure fluctuations in the mid-chord region since a smaller clearance \textcolor{black}{intensifies} the $TLF$ further upstream near the leading-edge \textcolor{black}{as shown by the PIV results. This increase in the turbulence near the leading-edge} is then also responsible for the concentrated region of high pressure fluctuations near the leading-edge for the smallest clearance observed in \cref{fig:Tip_Spectral_Contours} (i). Although the measurement range of the pressure fluctuations is limited to 10 kHz, it is clear that a lowering of clearance height enhances the high-frequency pressure fluctuations near the leading-edge which is consistent with this region being responsible for an amplification in the high-frequency far-field sound as observed in \cref{fig:DAMAS_Spectra_Tip_Gap_Comparison} (d). \par

\begin{figure}
\centering
  \centerline{\includegraphics[width = \textwidth]{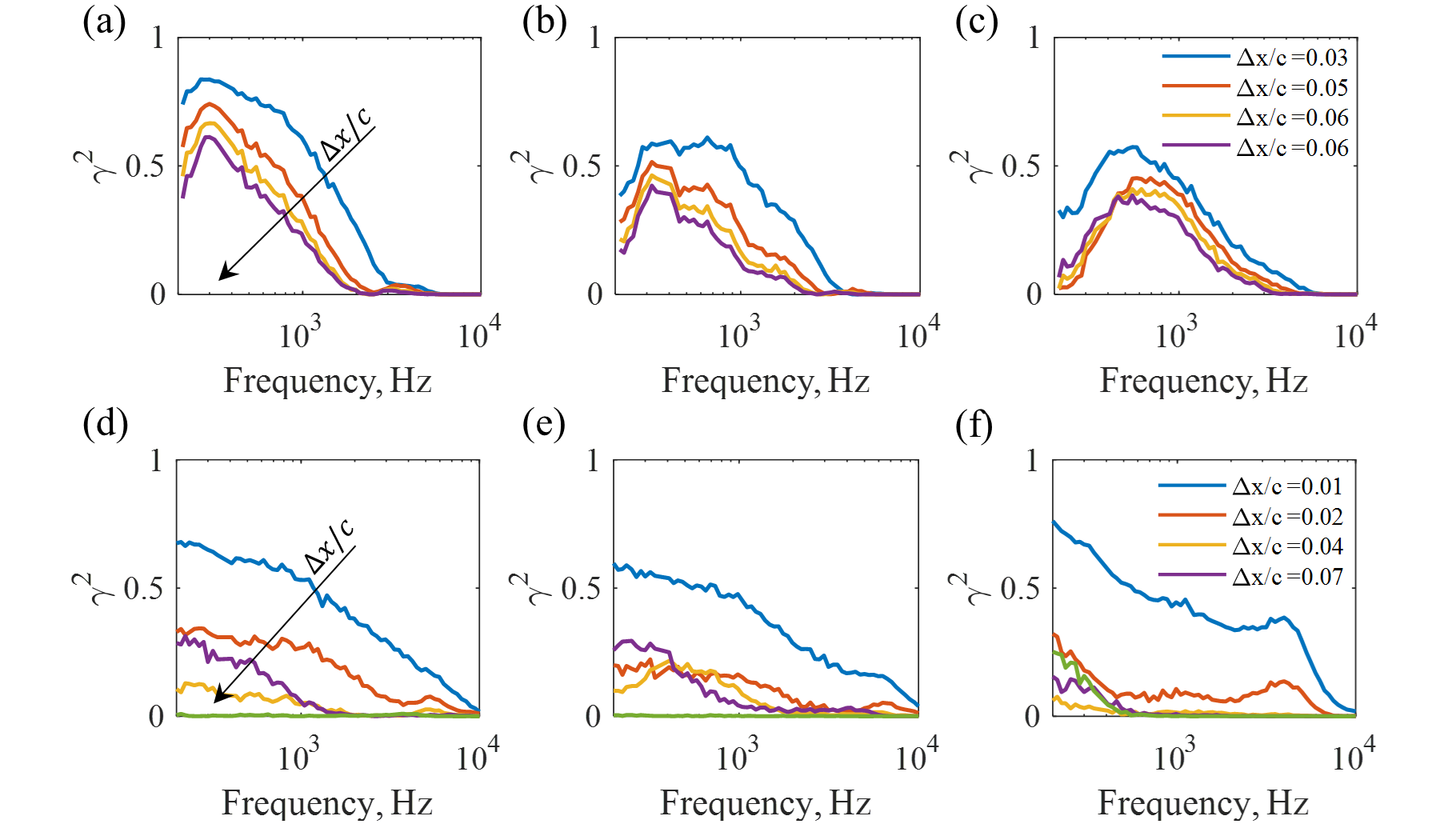}}
  \caption{Two-point statistics of tip surface pressure fluctuations. Coherence between the remote microphone at $x/c$ = 0.85 and the rest of the downstream microphones at $\alpha_g$ = 12.5$^\circ$ and $U_\infty$ = 30 m/s for $h$ = (a) 5.28 mm, (b) 3.96 mm and (c) 2.64 mm (legend shown in (c)). Coherence between the most upstream microphone at $x/c$ = 0.02 and five microphones downstream at $\alpha_g$ = 12.5$^\circ$ and $U_\infty$ = 30 m/s for $h$ = (d) 5.28 mm, (e) 3.96 mm and (f) 2.64 mm (legend shown in (f).}
\label{fig:Coherence}
\end{figure}

Finally, we briefly consider some two point statistics of surface pressure fluctuations on the tip surface to further understand the sound source behaviour in clearance flows. \Cref{fig:Coherence} (a) -- (c) show the coherence between pressure fluctuations on the tip near the trailing-edge for the three clearances at the highest angle of attack. The coherence was calculated between the remote microphone at $x/c$ = 0.85 and the remaining microphones downstream, see \cref{tab:Surf_Pressure_Locations} for measurement locations. These plots show that the coherence with statistical significance is limited to frequencies below 3 -- 4 kHz which implies the presence of large structures in the trailing-edge region and it is these structures which are responsible for the low-frequency noise generation observed earlier. As the clearance height is reduced, the low-frequency pressure fluctuations become less coherent which manifests as lower sound levels for these clearances in the far-field. \textcolor{black}{The behaviour of the pressure fluctuations near the trailing-edge is consistent with the PIV results discussed earlier (see \cref{fig:Instantaneous_Velocity}) that show the presence of large coherent structures in this region which become weaker as the clearance height is reduced.} \par

In contrast to the two-point statistics near the trailing-edge, the coherence function near the leading-edge (\cref{fig:Coherence} (d) -- (f)) has a different appearance. The coherence in the leading-edge region was calculated between the most upstream microphone ($x/c$ = 0.02) and five microphones downstream of it. In the leading-edge region, the coherence is not limited to the lowest frequencies and the shape of the coherence function is a stronger function of clearance height. For the two larger clearances (\cref{fig:Coherence} (d) and (e)), the coherence is similar with the exception that the reduction of coherence with spatial separation is slightly greater for smaller clearances. As the clearance height is reduced further (\cref{fig:Coherence} (f)), a decay in coherence and a local maximum in the coherence function around 4\,kHz is observed. These results indicate that in the leading-edge region a reduction in clearance height may serve to decrease the coherent length-scale of pressure fluctuations, but it also intensifies the amplitude of high-frequency pressure fluctuations \textcolor{black}{which results in larger high-frequency noise levels from this region. As shown by the PIV results, the interaction of the oncoming flow with the pressure-side edge of the tip for smaller clearances is likely the source of these high-frequency pressure fluctuations observed in both the near and the far-field.} \par

\pagebreak

\section{Conclusion}

Measurements of the far-field sound\textcolor{black}{, flow-field} and fluctuating surface pressure on the tip of a cambered airfoil were performed in a tip clearance flow formed by introducing a gap between the airfoil tip and a stationary, flat wall. The effect of Mach number, angle of attack and clearance height on the near and far-field statistics was explored. The tip clearance heights ($h$) considered ranged from approximately 14\% to 30\% of the incoming, undisturbed boundary-layer height on the adjacent flat-wall and for each clearance three geometric angles of attack ($\alpha_g$) equal to 5.6$^\circ$, 9.1$^\circ$, and 12.5$^\circ$ were considered. The Mach number ($M_\infty$) range in the measurements was 0.04 -- 0.13, while the Reynolds numbers based on clearance height ranged from 2.600 -- 16,000. \par

The character of the tip clearance noise and its dependence on the angle of attack has been clarified through a detailed analysis of beamformed sound source maps, \textcolor{black}{PIV datasets} and pressure fluctuations on the tip surface. \textcolor{black}{Although the experimental setup utilises an idealised configuration with a stationary airfoil and the exact flow conditions in turbomachinery applications with high turning angles may vary, the present work provides insight into the noise generated by typical flow features found in these applications.} It was observed that the tip clearance far-field sound across the measurable frequency range exceeds the noise generated by other sources (airfoil trailing-edge noise and airfoil-wall junction noise) for each configuration considered in the present work. At the lowest angle of attack considered, the dominant noise source resides close to the trailing-edge, but as the angle of attack increases the sound from the mid-chord and leading-edge regions of the clearance also becomes important. \textcolor{black}{This behaviour is consistent with the PIV results which show a significant increase in the cross-flow intensity with angle of attack across the mid-chord and leading-edge regions. It was observed that an increase in the cross-flow intensity in the mid-chord and leading-edge regions} mostly affects the mid-to-high frequency noise generation, while the low-frequency noise was controlled by a separate mechanism near the trailing-edge which is a stronger function of clearance height than the angle of attack. \textcolor{black}{PIV measurements reveal that this low-frequency noise source may be associated with the presence of a tip separation vortex system near the trailing-edge and the suppression of these vortices with a reduction in clearance height is responsible for a decrease in the low-frequency noise radiation levels for smaller clearances. The different character of the low frequency clearance noise was confirmed by the fact that it scaled with $M^5$ and Strouhal number based on free-stream velocity and clearance height. On the other hand, the mid-to-high frequency noise magnitude was found to scale on the sixth power of Mach number and Helmhotz number which implies the presence of significant acoustic scattering from the sharp tip edges and the adjacent wall. The fluctuating pressures measured on the tip were found to be consistent with the far-field sound results in that while an increase in the strength of the cross-flow with angle of attack amplifies the pressure fluctuations in the leading-edge and the mid-chord regions of the clearance significantly, those closer to the trailing-edge are only marginally affected. The results also show that source region near the trailing-edge consists of large structures which are the reason why this region remains a significant radiator of low-frequency far-field sound.} \par    

The effects of changing the clearance height on the far-field sound and the sound source region were also examined. The results show that while a reduction in clearance height decreases and increases sound levels at low and high frequencies, respectively, those in the mid-frequency range (3 -- 8 kHz) are affected only marginally. Although the amount of reduction and increase in noise with tip clearance varies, the trend remains the same for all three angles of attack considered. It is shown that the low-to-mid frequency noise reduction as the clearance height is decreased is driven by a reduction in the surface pressure fluctuations on the tip in both the mid-chord and the trailing-edge regions of the clearance, \textcolor{black}{possibly brought on by a reduction in the turbulence associated with the $TSV$}. The mid-chord region noise source weakens due to a weaker cross-flow in this region for smaller clearances, while the reduction in the trailing-edge source strength with clearance height is related to a reduction in the magnitude of the coherent surface pressure fluctuations \textcolor{black}{and turbulence intensity} in this region. On the other hand, the increase in high-frequency noise with lowering of the clearance height is associated with an amplification of the high-frequency pressure fluctuations near the leading-edge of the clearance \textcolor{black}{which was found to be a result of stronger interaction between the oncoming flow and the sharp edges of the tip}. \par

\section*{Acknowledgments}
The authors would like to thank the Defence Science and Technology Group (DSTG) for the financial support of this work.
\pagebreak

\bibliographystyle{unsrt}
\bibliography{References}

\begin{thebibliography}{10}

\bibitem{RN259}
B.~Lakshminarayana.
\newblock Methods of predicting the tip clearance effects in axial flow
  turbomachinery.
\newblock {\em Journal of Basic Engineering}, 92(3):467--480, 1970.

\bibitem{RN260}
M.~Inoue, M.~Kuroumaru, and M.~Fukuhara.
\newblock Behaviour of tip leakage flow behind and axial compressor rotor.
\newblock {\em Journal of Engineering for Gas Turbines and Power}, 108:7--14,
  1986.

\bibitem{RN261}
M.~Inoue and M.~Kuroumaru.
\newblock Structure of tip clearance flow in an isolated axial compressor
  rotor.
\newblock {\em Journal of Turbomachinery}, 111(3):250--256, 1989.

\bibitem{RN262}
A.~Yamamoto.
\newblock Endwall flow/loss mechanisms in a linear turbine cascade with blade
  tip clearance.
\newblock {\em Journal of Turbomachinery}, 111(3):264--275, 1989.

\bibitem{RN258}
B.~Lakshminarayana, M.~Zaccaria, and B.~Marathe.
\newblock The structure of tip clearance flow in axial flow compressors.
\newblock {\em Journal of Turbomachinery}, 117(3):336--347, 1995.

\bibitem{RN255}
C.~Muthanna and W.J. Devenport.
\newblock Wake of a compressor cascade with tip gap, part 1: Mean flow and
  turbulence structure.
\newblock {\em AIAA Journal}, 42(11):2320--2331, 2004.

\bibitem{RN265}
Q.~Tian and R.~Simpson.
\newblock Experimental study of tip leakage flow in the linear compressor
  cascade: Part i - stationary wall.
\newblock In {\em 45th AIAA Aerospace Sciences Meeting and Exhibit}, 2007.

\bibitem{RN256}
Y.~Wang and W.J. Devenport.
\newblock Wake of a compressor cascade with tip gap, part 2: Effects of endwall
  motion.
\newblock {\em AIAA Journal}, 42(11):2332--2340, 2004.

\bibitem{RN266}
Q.~Tian and R.~Simpson.
\newblock Experimental study of tip leakage flow in the linear compressor
  cascade: Part ii - effect of moving wall.
\newblock In {\em 45th AIAA Aerospace Sciences Meeting and Exhibit}, 2007.

\bibitem{RN264}
D.~You, R.~Mittal, M.~Wang, and P.~Moin.
\newblock Study of rotor tip-clearance flow using large-eddy simulation.
\newblock In {\em 41st Aerospace Sciences Meeting and Exhibit}, 2003.

\bibitem{RN268}
F.~Kameier and W.~Neise.
\newblock Experimental study of tip clearance losses and noise in axial
  turbomachines and their reduction.
\newblock {\em Journal of Turbomachinery}, 119:460--471, 1997.

\bibitem{RN270}
U.W. Ganz, P.D. Joppa, T.J. Patten, and D.F. Scharpf.
\newblock Boeing 18-inch fan rig broadband noise test.
\newblock Report NASA CR-1998-208704, NASA, 1998.

\bibitem{RN269}
M.R. Khorrami, F.~Li, and M.~Choudhari.
\newblock Novel approach for reducing rotor tip-clearance-induced noise in
  turbofan engines.
\newblock {\em AIAA Journal}, 40(8):1518--1528, 2002.

\bibitem{RN252}
M.C. Jacob, J.~Grilliat, R.~Camussi, and G.C. Gennaro.
\newblock Aeroacoustic investigation of a single airfoil tip leakage flow.
\newblock {\em International Journal of Aeroacoustics}, 9(3):253--272, 2010.

\bibitem{RN253}
R.~Koch, M.~Sanjosé, and S.~Moreau.
\newblock Large-eddy simulation of a single airfoil tip-leakage flow.
\newblock {\em AIAA Journal}, 59(7):2546--2557, 2021.

\bibitem{RN251}
S.~Palleja-Cabre, B.J. Tester, R.~Jeremy Astley, and G.~Bampanis.
\newblock Aeroacoustic assessment of performance of overtip liners in reducing
  airfoil noise.
\newblock {\em AIAA Journal}, pages 1--16, 2021.

\bibitem{RN254}
J.~Boudet, M.C. Jacob, J.~Caro, E.~Jondeau, and B.~Li.
\newblock Wavelet analysis of a blade tip-leakage flow.
\newblock {\em AIAA Journal}, 56(8):3332--3336, 2018.

\bibitem{RN189}
J.~Grilliat, M.~Jacob, R.~Camussi, and G.~Caputi-Gennaro.
\newblock Tip leakage experiment - part one: Aerodynamic and acoustic
  measurements.
\newblock In {\em 13th AIAA/CEAS Aeroacoustics Conference (28th AIAA
  Aeroacoustics Conference)}, pages AIAA 2007--3684. AIAA, 2007.

\bibitem{RN272}
R.~Camussi, G.~Caputi Gennaro, M.~Jacob, and J.~Grilliat.
\newblock Experimental study of a tip leakage flow - part two: Wavelet analysis
  of wall pressure fluctuations.
\newblock In {\em 13th AIAA/CEAS Aeroacoustics Conference (28th AIAA
  Aeroacoustics Conference)}, 2007.

\bibitem{RN273}
R.~Camussi, J.~Grilliat, G.~Caputi-Gennaro, and M.C. Jacob.
\newblock Experimental study of a tip leakage flow: wavelet analysis of
  pressure fluctuations.
\newblock {\em Journal of Fluid Mechanics}, 660:87--113, 2010.

\bibitem{RN267}
T.~Fukano, Y.~Takamatsu, and Y.~Kodama.
\newblock The effects of tip clearance on the noise of low pressure axial and
  mixed flow fans.
\newblock {\em Journal of Sound and Vibration}, 105(2):291--308, 1986.

\bibitem{RN274}
J.~Grilliat, M.~Jacob, E.~Jondeau, M.~Roger, and R.~Camussi.
\newblock Broaband noise prediction models and measurements of tip leakage
  flows.
\newblock In {\em 14th AIAA/CEAS Aeroacoustics Conference (29th AIAA
  Aeroacoustics Conference)}, 2008.

\bibitem{RN334}
D.~Moreau, C.d. Silva, R.~Kisler, J.~Tan, C.~Jiang, M.~Awasthi, and C.~Doolan.
\newblock The design and characterisation of the unsw anechoic wind tunnel.
\newblock In {\em 23rd Australasian Fluid Mechanics Conference}, Sydney,
  Australia, 2022.

\bibitem{RN16}
D.J. Moreau, Z.~Prime, R.~Porteous, C.J. Doolan, and V.~Valeau.
\newblock Flow-induced noise of a wall-mounted finite airfoil at
  low-to-moderate reynolds number.
\newblock {\em Journal of Sound and Vibration}, 333:6924–6941, 2014.

\bibitem{RN17}
D.J. Moreau and C.J. Doolan.
\newblock Tonal noise production from a wall-mounted finite airfoil.
\newblock {\em Journal of Sound and Vibration}, 363:199--224, 2016.

\bibitem{RN111}
T.J. Mueller.
\newblock {\em Aeroacoustic Measurements}.
\newblock Experimental Fluid Mechanics. Springer-Verlag Berlin Heidelberg,
  2002.

\bibitem{RN230}
E.~Sarradj.
\newblock Three-dimensional acoustic source mapping with different beamforming
  steering vector formulations.
\newblock {\em Advances in Acoustics and Vibration}, 2012:292695, 2012.

\bibitem{RN37}
M.~Awasthi.
\newblock {\em Sound Radiated from Turbulent Flow over Two and
  Three-Dimensional Surface Discontinuities}.
\newblock Doctoral dissertation, 2015.

\bibitem{RN231}
T.~Padois, C.~Prax, and V.~Valeau.
\newblock Numerical validation of shear flow corrections for beamforming
  acoustic source localisation in open wind-tunnels.
\newblock {\em Applied Acoustics}, 74(4):591--601, 2013.

\bibitem{RN224}
T.F. Brooks and W.M. Humphreys.
\newblock A deconvolution approach for the mapping of acoustic sources (damas)
  determined from phased microphone arrays.
\newblock {\em Journal of Sound and Vibration}, 294(4-5):856--879, 2006.

\bibitem{RN229}
R.~Merino-Martínez, P.~Sijtsma, A.R. Carpio, R.~Zamponi, S.~Luesutthiviboon,
  A.M.N. Malgoezar, M.~Snellen, C.~Schram, and D.G. Simons.
\newblock Integration methods for distributed sound sources.
\newblock {\em International Journal of Aeroacoustics}, 18(4-5):444--469, 2019.

\bibitem{RN204}
M.~Awasthi, J.~Rowlands, D.J. Moreau, and C.J. Doolan.
\newblock Two-step hybrid calibration of remote microphones.
\newblock {\em The Journal of the Acoustical Society of America}, 144, 2018.

\end{thebibliography}

\end{document}